\documentclass{article}

\usepackage{geometry}
\usepackage{graphicx}
\usepackage{newtxtext}
\usepackage{newtxmath}
\usepackage{natbib}
\usepackage{hyperref}

\usepackage{amsmath}

\usepackage[utf8]{inputenc}

\usepackage{xcolor}
\usepackage[ruled,vlined]{algorithm2e}

\SetCommentSty{mycommfont}
\usepackage{bm}
\usepackage{cases}
\graphicspath{{././}}
\usepackage[section]{placeins}
\usepackage{siunitx}

\hypersetup{
    colorlinks = true,
    urlcolor   = blue,
    citecolor  = black,
}

\newcommand{\RomanNumeralCaps}[1]
\linenumbers

\newcommand{\com}[1]{{\color{black} #1}}

\newcommand{\vcon}{{\bf{v}}}
\newcommand{\vmes}{{\bf v}'}
\newcommand{\obc}{\Omega_{\text{bc}}'}
\newcommand{\omes}{\Omega'}
\newcommand{\ocon}{{\Omega}}

\newcommand{\obuf}{\Omega_0}

\newcommand{\pima}{\bar{\bm \pi}}

\newcommand{\kbt}{k_BT}
\newcommand{\fij}{F_{ij}}
\newcommand{\rij}{r_{ij}}
\newcommand{\rbij}{{\bf r}_{ij}}

\newcommand{\fijcon}{{F}_{IJ}}
\newcommand{\rijcon}{{r}_{IJ}}
\newcommand{\rbijcon}{{\bf r}_{IJ}}
\newcommand{\ebijcon}{{\bf e}_{IJ}}

\newcommand{\facon}{a_{IJ}}
\newcommand{\fbcon}{b_{IJ}}

\newcommand{\fijmes}{{F}_{ij}'}

\newcommand{\rbijmes}{{\bf r}_{ij}'}
\newcommand{\ebijmes}{{\bf e}_{ij}'}

\newcommand{\Aij}{A_{ij}}
\newcommand{\Bij}{B_{ij}}

\newcommand{\edot}{\dot{\varepsilon}}
\newcommand{\gdot}{\dot{\gamma}_x}
\newcommand{\gdoty}{\dot{\gamma}_y}

\newcommand{\vb}{{\bf v}}

\newcommand{\vbtilde}{{\bf\tilde{v}}}
\newcommand{\wbbarij}{{\bf \tilde{W}}_{ij}}
\newcommand{\wbij}{{\bf {W}}_{ij}}
\newcommand{\fa}{a_{ij}}
\newcommand{\fb}{b_{ij}}

\newcommand{\der}{\text{d}}

\newcommand{\dtmic}{\Delta {t}'}
\newcommand{\dtmac}{\Delta {t}}

\newcommand{\lammac}{{\lambda}}
\newcommand{\lammic}{{\lambda}'}

\newcommand{\dhmi}{\text{d}x'}
\newcommand{\dhma}{\text{d}{x}}

\newcommand{\sdpd}{smoothed dissipative particle dynamics}
\makeatletter							% Begin definition
\def\printtitle{%						% Define command: \printtitle
	{\centering \huge \sc \textbf{\@title}\par}}		% Typesetting
\makeatother

\makeatletter							% Begin definition
\def\printauthor{%					% Define command: \printauthor
	{\centering \small \@author}}				% Typesetting
\makeatother

\title{Generalized Lagrangian Heterogenous Multiscale Modeling of Complex Fluids}

\author{%
Nicolas Moreno$^{1}*$  and Marco Ellero$^{1,2,3} \dagger$ \\
{\footnotesize 1.Basque Center for Applied Mathematics (BCAM), Alameda de Mazarredo 14, Bilbao 48400, Spain\\
\footnotesize 2. IKERBASQUE, Basque Foundation for Science, Calle de Maria Dias de Haro 3, 48013, Bilbao, Spain\\
\footnotesize 3. Zienkiewicz Center for Computational Engineering (ZCCE), Swansea University, Bay Campus, Swansea SA1 8EN, United Kingdom}\\
 {$*$ nmoreno@bcamath.org}  -- {$\dagger$ mellero@bcamath.org}\\
}
\date{2022}
\begin{document}
\printtitle 
	
	\printauthor

\begin{abstract}
We introduce a  full-Lagrangian heterogeneous multiscale method (LHMM) to model complex fluids with microscopic features that can extend over large spatio/temporal scales, such as polymeric solutions and multiphasic systems. The proposed approach discretizes the fluctuating Navier-Stokes equations in a particle-based setting using Smoothed Dissipative Particle Dynamics (SDPD). This multiscale method uses microscopic information derived on-the-fly to provide the stress tensor of the momentum balance in a macroscale problem, therefore bypassing the need for approximate constitutive relations for the stress. We exploit the intrinsic multiscale features of SDPD to account for thermal fluctuations as the characteristic size of the discretizing particles decrease. We validate the LHMM using different flow configurations (reverse Poiseuille flow, flow passing a cylinder array, and flow around a  square cavity) and fluid (Newtonian and non-Newtonian). We showed the  framework's flexibility to model complex fluids at the microscale using multiphase and polymeric systems. We showed that stresses are adequately captured and passed from micro to macro scales, leading to richer fluid response at the continuum. In general, the proposed methodology provides a natural link between variations at a  macroscale, whereas accounting for memory effects of microscales.
\end{abstract}

% \begin{keywords}
% Authors should not enter keywords on the manuscript, as these must be chosen by the author during the online submission process and will then be added during the typesetting process (see \href{https://www.cambridge.org/core/journals/journal-of-fluid-mechanics/information/list-of-keywords}{Keyword PDF} for the full list).  Other classifications will be added at the same time.
% \end{keywords}
% 
% {\bf MSC Codes }  {\it(Optional)} Please enter your MSC Codes here

\section{Introduction}
\begin{figure}%[!tbhp]%tbhp]
 	\centering
	\setlength{\unitlength}{0.1\columnwidth}
	\scalebox{0.8}{
	\includegraphics[trim=0cm 0cm 0cm 0cm, clip=true, width=1\textwidth]{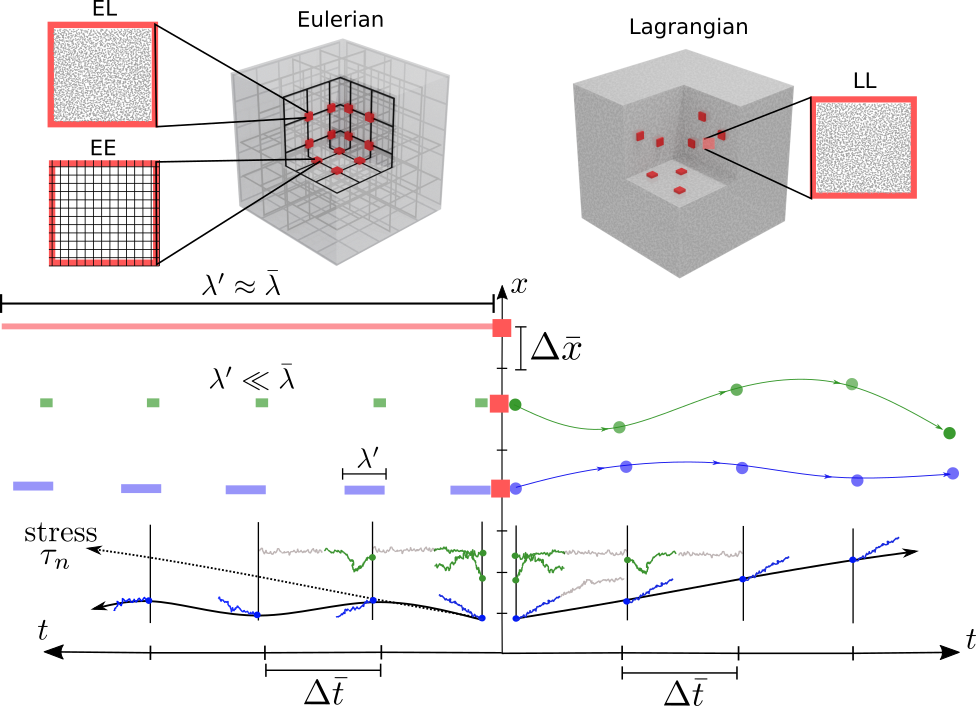}
	}
	\caption{Scheme of different HMM approaches. Eulerian - Eulerian (EE), Eulerian-Lagrangian (EL), and Lagrangian-Lagrangian (LL). The evolution of the stress tensor depends on the effective relaxation times at the microscales $\lambda'$. Systems with $\lambda' \ll \bar{\lambda}$ (green) are accurately computed at the microscopic scales, whereas for $\lambda' \leq \bar{\lambda}$ (blue) larger microscale simulations are required to capture memory effects as the macro scales evolve. LL approaches facilitate the carrying of the stress information during the time integration at macroscales.}
	\label{fig:hmms}
\end{figure}

The modelling of complex fluids, synthetic or biological, is in general a challenging task due to the multiscale nature of the flow, leading to complex behaviours such as flow-induced phase separation, shear-thinning/thickening, and viscoelasticity.  Usual approaches involve the solution of a macroscopic balance of momentum, along with constitutive equations that relate the dependency of the stresses and velocity fields due to microscopically-originated features. However, limitations of these approaches arise when the constitutive equations are not known \textit{a priori}. Moreover, the existence of large relaxation times at the microscale originates a non-trivial interplay with macroscopic flow features, requiring a detailed description of the entire stress history. In this context, heterogeneous multiscale methods (HMM)\citep{Engquist2007} that combine numerical algorithms to resolve separately macro and micro-scales, appear as powerful tools to model the behaviour of fluids across scales.  In HMMs, microscales are localized and solved on parts of the domain to obtain microscopically-derived properties that are used to close the macroscale problem\citep{Ren2005}. This methodology offers the advantage of capturing microscopic effects at the macroscopic length scales, with a lower cost than solving the full microscale problem in the whole domain. In HMMs the derived microscales properties can enter into the macroscales representations either through constitutive relationships, or microscopic stresses information without \textit{a priori} assumption of the constitutive relationships. The latest is an important advantage of HMMs for the modelling of complex fluids. For an extended review on HMMs, the reader is referred to \citep{Engquist2007}.
% % appear as powerfull tools for capturing the behavior of fluids across scales. 
% % Polymeric fluids represent stress-strain and stress-strain rate historical dependency coming from microscopic polymer dynamics. 
% To account for the effects of fluid and flow complexity it is customary in HMMs to define the viscous stress tensor in terms of microscopic properties of the system, which can be incorporated through constitutive equations or directly estimated solving the a microscopic problem\citep{example}.
\\
\\ 
Depending on the type of discretization (Eulerian or Lagrangian) used for macro and micro scales, the HMMs are classified as Eulerian/Eulerian (EE), Lagrangian/Lagrangian (LL), Eulerian/Lagrangian (EL), and Lagrangian/Eulerian (LE). See figure \ref{fig:hmms}$.a$. A large part of the existent HMMs relies on EE and EL schemes\citep{Engquist2007}, where the macroscale dynamics are resolved on a fixed grid (using a variety of methods such as finite elements, finite volumes, lattice Boltzmann, to name a few), and microscale simulations (e.g. molecular dynamics\citep{Alexiadis2013,Borg2015,Tedeschi2021}, coarse-graining methods, stochastic methods, etc) are associated to grid points, where microscopic properties are derived. For viscoelastic fluids modelling, Laso and \"Ottinger introduced a pioneering approach known as CONNFFESSIT\citep{Laso1993} (Calculation of Non-Newtonian Flow: Finite Element and Stochastic Simulation Technique), combining finite elements at the macroscale and stochastic particle simulations of polymer dynamics at the microscale.
\\
\\
EE and EL approaches are in general suitable for fluids with microstructural relaxation times ($\lammic$) sufficiently small compared to the macroscopic ones ($\lammac$) \citep{Ren2005,Yasuda2008,Yasuda2014}. As depicted in figure \ref{fig:hmms}$.b$, for multiscale problems with a large time scale separation, $\lammic \ll \lammac$, an equilibrated response of the microscopic stresses can be obtained in relatively short intervals, regardless of the flow history, since for practical purposes the microscales are seen by the macro solvers as quasi-steady solutions, independently of their initial configuration. Such strategy has been applied to atomistic-continuum simulations of simple fluids (Ren and E (2005)) using Molecular Dynamics with an Eulerian grid-based calculation of the flow field. This can be easily done in simple fluids where the local stress depends point-wise in time on the velocity gradient. Thus, the initial conditions for the microstructure (atoms positions/velocities) can be chosen arbitrarily at every time step and the average stress is calculated via the Irving-Kirkwood approximation, provided that local stationarity is achieved within the same time step.  The previous approach, however, cannot be applied to complex fluids with finite memory, where stresses (and microstructure) do heavily depend on flow history and relaxation times are likely to be comparable or largely exceed the macroscopic time step. Using directly EL or EE schemes, it is fundamentally and technically restrictive to generate such an initial microstructural configuration in a fixed fluid cell for at least two reasons: \textit{1)} it is a priori not known where the fluid comes from and what its flow history was, and \textit{2)} even if this sequential macroscopic information would be accessible in a given element of fluid, it requires additional constitutive and numerical features able to account for complex spatio/temporal variations. Alternatives to address the issue \textit{1)} include spatio/temporal homogenization techniques, backward-tracking Lagrangian particles combined with Eulerian grids to capture memory effects in the fluid\citep{Phillips1999, Wapperom2000, Ingelsten2021}. However, as already mentioned, fluid memory can be very long in polymer systems, suspensions, etc. precluding simple linear backward approximations. Regarding the issue  \textit{2)}, one alternative is to incorporate continuum configuration fields\citep{Ottinger1997} that can be discretized and advected from the macroscales. Nevertheless, in this case, it would be extremely difficult to know a priori those fields for general multiphysics problems (i.e. non-polymeric), as well as the numerical generation of microscopic configuration consistent with the history of the fluid.
% From a computational standpoint, EE and EL can be applied directly to appoximate the such memory effect, however it will require larger time integrations at the microscale. Reaching a computationally-prohibitive condition for $\lammic \approx \lammac$, where the microscopic simulations need to be conducted over macroscopic times scales. 

For systems with larger microstructural relaxation times, the particular restrictions of EE and EL can be circumvented using fully Lagrangian, LL, schemes\citep{Engquist2007}, and a proper sampling procedure for the microstructure. Indeed, LL schemes have been successfully used to model elastic effect and history-dependent flows \citep{Murashima2010,Seryo2020,Morii2020}. As illustrated in figure \ref{fig:hmms}$.b$, LL schemes directly track the material points at the macroscale retaining their strain and strain-rate variation, thus naturally handling history-dependent fluids. A variety of LL methodologies have emerged over the last decade, adopting mainly smoothed particle hydrodynamics (SPH) discretizations at the macro scales and combination of different microscopic models \citep{Ellero2003,Murashima2010,Xu2016,Feng2016,Sato2017, Zhao2018, Sato2019, Seryo2020,Morii2020,Schieber2020,VanDerGiessen2020}. At the microscales, the stress evolution of polymeric solutions and entanglements have been accounted for using Brownian dynamic\citep{Xu2016}, active learning\citep{Zhao2018, Seryo2020}, and slip-link models\citep{Feng2016, Sato2017, Sato2019}. In these LL schemes, it is considered that micro scales only account for the polymer contribution to the stress, whereas fluid is modelled uniquely from the macroscopic discretization\citep{Feng2016}. Its effect (i.e. velocity gradient tensor) enters the Langevin-type dynamics for stochastic micro-realizations implicitly as a single parameter, and not directly as a boundary condition for the full micro-system.
\\
\\
% slip-link models to derive rheological properties\cite{Becerra2020}
% % At the microscale, the definition of boundary conditions has typically posed some restrictions for LL methods. Due to the history-dependent evolution of the flow and the existence of non-trivial flow configurations, the microscales can be subjected to arbitrary series of deformations that are usually difficult to handle with traditional periodic boundary conditions.
In fact, one important issue limiting the applicability of HMM methods to more detailed descriptions of complex fluids is precisely the proper imposition of microscale constraints that are consistent with the macroscale kinematics and the calculation of microscopic information required by the macro state\citep{Engquist2007}. When using particle-based micro-models with explicit solvent description (e.g. MD, DPD, DEM, SDPD), the construction of this constrained microscale solver represents often the most cumbersome technical step. For LL schemes, due to the history-dependent evolution of the flow and the existence of non-trivial flow configurations, the microscales can be subjected to arbitrary series of deformations that are usually difficult to handle with traditional periodic boundary conditions (BCs). To avoid these limitations, existent LL schemes have been restricted to the use of microscopic simulators that do not dependent on the \textit{``physical''} boundary conditions\citep{Feng2016, Sato2017, Sato2019,Morii2020}. This include, for example, the case of BD for statistically independent polymers, such as dilute polymer solutions or polymer melts in mean field approximation, or that utilize geometries that reproduce simple flow configurations\citep{Seryo2020} (i.e. simple shear or uniaxial deformation). More general micro-macro couplings (e.g. full particle-based model of polymeric dispersions, colloid suspensions, emulsions, etc.) involving detailed micro-systems models undergoing arbitrarily flow deformations are beyond the capabilities of the current frameworks.
% heat and momentum transfer occurs in macroscales via finite volume method and NEMD in micro in simple shear flow\cite{Yasuda2014}
% RoadMap for multiscale modelling of materials\cite{VanDerGiessen2020}
% Two scale model that cannot be used with CONNFFESSIT to solve numerically\cite{Semkiv2017}, thus requiring other alternative methods.
\\
\\
Moreover, for micro solvers that adopt mean-field approximations, one important assumption is that the microscopic states of all polymers are in equilibrium and that the coils do not have translational degrees of freedom, but only rotational and extensional ones\citep{Morii2020}. Regarding microscopic BCs approaches using simple flow configurations, they are suitable to account for translational effects and often provide information sufficient to characterize simple fluids. However, since complex fluids can possess microscopic structures that are influenced by different flow configurations, geometries, time scales, and deformation rates, it has been evidenced that to correctly model non-Newtonian fluids\citep{Tedeschi2021}, it is necessary to determine the full stress contribution from the microscopic solver. 
\\
\\
In this manuscript, we propose a generalized fully Lagrangian HMM (LHMM) using \sdpd \citep{Espanol2003a, Ellero2018} (SDPD), suitable to model general complex fluids (e.g. colloids, polymer, microstructures in suspensions) while using the same fluid description across scales. Among the different computational methods successfully used to model Newtonian and non-Newtonian fluids at continuum and microscales, SDPD has emerged as a suitable tool to simulate complex fluids \citep{Kulkarni2013, Muller2014, Ellero2018}. The main strengths of SDPD are \textit{i)} it consistently discretizes the fluctuating Navier-Stokes equations allowing the direct specification of transport properties such as viscosity of the fluid; \textit{ii)} SDPD is compliant with the General Equation for Nonequilibrium Reversible-Irreversible Coupling (GENERIC)\citep{Ottinger2005}, and therefore, it discretely satisfies the First and Second Laws of Thermodynamics, and Fluctuation-Dissipation Theorem (FDT); \textit{iii)} at macroscopic scales SDPD converges to the well-known continuum method smoothed particle hydrodynamics (SPH) as the characteristic size of the discretized particle increases \citep{Vazquez-Quesada2009, Ellero2018}. For an extended review of SDPD, the reader is referred to the publication of Ellero and Espa\~nol\citep{Ellero2018}.
\\
\\
Since SDPD offers a natural physical link between different scales, we construct an HMM that uses SDPD to solve both macro and microscales. This approach ensures the compatibility of the different representations by construction and physical consistency across scales. At the microscales, we adopt the recently proposed BC methodology \citep{Moreno2021} that allows the acquisition of the full microscopic stress contributions for \textit{arbitrary} flow configurations. This allows to carry out micro-computations under general mixed flow conditions. Furthermore, compared to existing LL methodologies, our approach exploits the versatility of SDPD to model a variety of microscopic physical systems beyond polymeric systems. We  can summarise the main features of the proposed LHHM framework as

\begin{itemize}
 \item Model history-dependent flows by construction.
 \item Significant spatio-temporal gains in simulations compared to fully resolved microscale simulations. 
 \item Thermodynamic-consistent discretization of the Navier-Stokes equations in both macro-micro scales (deterministic - stochastic) providing a direct link to physical parameters.
 \item GENERIC compliant at both macro and micro levels.
 \item Multiphysics -- polymers, colloids, suspensions, multiphase systems --. No constitutive models for closure are required.
 \item Complex-flow configurations are allowed and can be handled it at the microscales.
\end{itemize}

% %
% % Eventhough simple flow configurations often provides information sufficient to characterize simple fluids. Recent investigations have evidenced that for complex fluids, it is necessary to determine the full stress contribution from microscales\citep{Tedeschi2021}. 
% %
% % Therefore, for complex fluids it is necessary that the stress contribution can be fully determined from microscales\citep{Tedeschi2021}.
% % Recent investigations combining finite elements and non-equilibrium molecular dynamics (NEMD) at microscale\citep{Tedeschi2021}, have shown that for complex  fluid it is necessary stress contribution can be fully determined from microscales
% % 
% % Simple shear is very well investigated for it is more easily realized in experiments than other flows. It
% 
% % provides fundamental information that is often sufficient to characterize simple fluids. However, complex
% 
% % fluids show a molecular structure that is able to change with respect to different conditions of motion,
% 
% % geometries, but also time scales and deformation rates
% %
% %However, since we investigate systems with an inherent length-scale separation, the increase of the microdomain in more than one order of magnitude in impractical.
% %\\
% %\\
% %\textit{PENDING: Comments of scale separation assumptions}
% \\
% \\
In the following sections, first, a general description of HMM is introduced along with the governing NS equations, then, the proposed fully Lagrangian approach and the particle-based discretization are presented. Finally, without loss of generality, we streamline the validation of the methodology focusing on two-dimensional simulations of complex flows with memory.  At the microscales, we adopt generic, yet complex, polymeric and multiphase flows to showcase the flexibility of the method. 

% In this manuscript we propose a fully Lagrangian HMM using particle-based discretizations for both scales. In this way, microscales simulations can performed on demand at arbitrary locations of the macroscale. Whereas the macroscales naturally carry the microscale flow history during the macroscopic time integration.

\section{Heterogeneous multiscale methods}
 
In general for HMMs, we can define the macroscopic problem considering a domain $\Omega \subset \mathcal{R}^D$ (with dimension $D=2,3$) with a boundary $\partial \Omega = \Gamma_{\mathcal{D}} \cup \Gamma_{\mathcal{N}}$, where $\Gamma_{\mathcal{D}}$ and $\Gamma_{\mathcal{N}}$ correspond to boundary regions where Dirichlet and Neumann boundary conditions are applied, respectively. The mass and momentum balance of the system in terms of the Navier-Stokes equations for an incompressible fluid with constant density $\rho$ can be expressed as
% %
% %\begin{align}
% %\begin{matrix}
% %\nabla \cdot \vb &= 0 &\text{in} \quad \Omega \times (0,T) \\
% %\rho\frac{\partial \vb}{\partial t} + \rho(\vb \cdot \nabla)\vb - \nabla \cdot \tau(\vb,p) &= f &\text{in} \quad \Omega \times (0,T) \\
% %\vb &= g &\text{on} \quad  \Gamma_D \times (0,T) \\
% %\tau(\vb,p){\bf\hat{n}} &= h &\text{on} \quad  \Gamma_N \times (0,T)\\
% %\vb(0) &= \vb_0 &\text{in} \quad \Omega \times \{0\}
% %\end{matrix}
% %\end{align}
% %
\begin{equation}
\begin{cases}{}
\nabla \cdot \vb &= 0 \quad \text{in} \quad \Omega \times (0,T), \\
\rho \frac{\der \vb}{\der t} - \nabla \cdot \bm\tau(\vb,p) &= f \quad \text{in} \quad  \Omega \times (0,T), \\
\vb &= g \quad \text{on} \quad  \Gamma_{\mathcal{D}} \times (0,T), \\
\vb(0) &= \vb_0 \quad  \text{in} \quad \Omega \times \{0\},
\end{cases}
\label{eq:bal}
\end{equation}
% %
% %The viscous stress tensor ${\bf \tau}$ is given by ${\bf \tau} = p{\bf I} + 2\eta \varepsilon$, where  $\varepsilon = \frac{1}{2}(\nabla \vb + (\nabla \vb)^{T})$.
where, the total stress tensor is given by ${\bm\tau} = p{\bf I} + \bm\pi$, being $p$ the pressure and $\bm \pi$ the viscous stress. For incompressible Newtonian fluids the viscous stress is a linear function of the strain rate ($\bm{\pi}=\eta(\nabla{\bf v} + \nabla{\bf v}^T)$, being $\eta$ the viscosity) and the flow can be totally described using \eqref{eq:bal}. For non-Newtonian fluids such as colloidal and polymeric systems, this linear relationship does not hold and constitutive equations are required\citep{Bird1987}. Additionally, for of microfluidics, where complex flow patterns and thermal effects may arise, the use  of Dirichlet boundary conditions, $\vb = g$ on $\Gamma_{\mathcal{D}} \times (0,T) $ may not accurately model such microscopic effects, requiring more elaborated considerations for the boundary conditions.
% To account for the effects of fluid and flow complexity it is customary in HMMs to define the viscous stress tensor in terms of microscopic properties of the system, which can be incorporated through constitutive equations or directly estimated solving the a microscopic problem\citep{example}.

% % The vast majority of the existent HMMs, adopt Eulerian approaches to solve the macroscopic problem \eqref{eq:bal} by discretizing $\Omega$ with fixed grid points\citep{}. The microscale (to retrieve the stress contributions) is solved at specific grid positions. The microscale problem is then solved using either Eulerian or Lagrangian discretizations, if microscales are explicitly modelled. 

\subsection{Lagrangian heterogeneous multiscale method (LHMM)}

We propose a LL-type of methodology, as depicted in figure \ref{fig:hmms}, discretizing both macro and micro scales with a particle-based representation of the system. We distinguish macroscale parameters and variables if they are derived from microscales calculations using the upper bar (i.e. $\bar{X}$), whereas microscale variables are denoted using a prime (i.e. $X'$). We use the subindex $x,y,$ and $z$, to indicate the coordinate axis. If we express the macroscopic viscous stress \textit{determined} from microscopic simulations in terms of hydrodynamic, non-hydrodynamic, and kinetic contributions as ${\bar{\bm \pi}}={\bar{\bm \pi}}^{h}+{\bar{\bm \pi}}^{*}+{\bar{\bm \pi}}^{k}$. The ensemble average stress can be represented by   
% $\bar{\bm \pi}({\bm v'}, t') = \bar{\bm \pi}^o({\bm v'}, t') + \bar{\bm \pi}^*({\bm v'}, t')$; 
% 
\begin{align}
%   \langle {\bar{\bm \pi}} \rangle&=\langle{\bar{\bm \pi}}^{h}\rangle+\langle{\bar{\bm \pi}}^{*}\rangle+\langle{\bar{\bm \pi}}^{k}\rangle \\
  \langle \bar{\bm \pi} \rangle &= \frac{1}{\Omega} \int_{\Omega} \bar{\bm \pi} \der \Omega, \nonumber \\
 &= \frac{1}{\Omega} \left(\int_{\Omega^h} \bar{\bm \pi}^h \der \Omega   + \int_{\Omega^*}  \bar{\bm \pi}^* \der \Omega + \int_{\Omega^k} \bar{\bm \pi}^k \der \Omega \right),
\label{eq:pimean}
 \end{align}
% 
% 
% \begin{align}
% c_{i} &=v_{i}-\left\langle v_{i}\right\rangle \\
% \langle\pi\rangle =2 \eta\left[\frac{1}{N} \sum_{i=1}^{N} \frac{1}{2} \sum_{j}\left(c_{i}-c_{j}\right) \otimes e_{i j} r_{j} w_{i j}^{\prime}+\sum_{j}\left[\left(c_{i}-c_{j}\right) \otimes e_{i j}\right]^{\top} r_{j} w_{i j}^{\prime}\right]
% \end{align}
% 
% \begin{align}
%  \langle \bar{\bm \pi} \rangle &= \frac{1}{\Omega} \int_{\Omega} \bar{\bm \pi} \der \Omega \\
%  &= \frac{1}{\Omega} \left(\int_{\Omega^h} \bar{\bm \pi}^h \der \Omega^h + \int_{\Omega^*} \bm \bar{\bm \pi}^* \der \Omega^* \right) 
% \end{align}
%
where $\bar{\bm \pi}^h$ accounts for the hydrodynamic contributions to the stress, and $\bar{\bm \pi}^*$ corresponds to the non-hydrodynamics effects (presence of colloids, polymers, walls, etc). In general, hydrodynamic contributions combine both ideal and non-ideal interactions, this is
\begin{equation}
\langle\bar{\bm \pi}^h\rangle=\langle\bar{\bm \pi}^o\rangle+\langle\bar{\bm \pi}^h|_{\text{non-ideal}}\rangle.
\label{eq:pihydro}
\end{equation}
Whereas the ideal effects are expected to occur in the fluid at all scales, the non-ideal interactions are only originated at microscales by the disruption of the flow field due to the presence of polymer, colloids, walls, or microstructures. Considering that the Newtonian (ideal) stress, in absence of complex microscopic effects is given by $\bar{\bm \pi}^o = 2\eta \bar{\textbf{d}}$ (being $\bar{\textbf{d}}$ the rate-of-strain tensor computed from microscopic information), we can rewrite \eqref{eq:pihydro} in the form 
\begin{align}
\langle{\bar{\bm \pi}}^{h}\rangle =\underbrace{\frac{1}{\Omega} \int_{\Omega} 2 \eta  \bar{\textbf{d}} \, \der \Omega}_{\text{ideal}}+\underbrace{\frac{1}{\Omega} \int_{\Omega} {\bar{\bm \pi}}^{h}(r)- 2\eta  \bar{\textbf{d}} \,\der \Omega}_{\text{non-ideal}}. \label{eq:decompose}
\end{align}
Now, if we consider that ideal stress contributes homogeneously over the whole domain a mean-field approximation holds and the ideal term of \eqref{eq:decompose} can be written in terms of macroscopic variables, ${\bm \pi}^o (\vcon, {t}) = 2 \eta  \textbf{d} =\langle \bar{\bm \pi}^o(\vmes, t')\rangle$. (Notice that the overbar notation of $\textbf{d}$ is omitted since is not a multiscale contribution. In contrast to $\bar{\textbf{d}}$ that is a macroscopic stress determined from microscopic variables). Thus, we can now introduce a hybrid macro-micro formulation of \eqref{eq:decompose} given by 
\begin{align}
\langle{\bar{\bm \pi}}^{h}\rangle = \underbrace{\epsilon 2 \eta  \textbf{d}}_{\text{macro}} + \underbrace{\frac{1}{\Omega} \int_{\Omega} {\bar{\bm \pi}}^{h}(r)- \epsilon 2\eta  \bar{\textbf{d}} \,\der \Omega}_{\text{micro}}.
\label{eq:micromacropi}
\end{align}
This scheme is a generalized framework that allows us to incorporate ideal hydrodynamics interactions of the fluid from both scales. The weighting parameter $\epsilon$ conveniently provides numerical stability to the method, whereas naturally accounting for spatial inhomogeneities of the stresses. According to \eqref{eq:micromacropi}, if $\epsilon=1$, the ideal hydrodynamic contributions are fully accounted for from the macroscale level, and microscales only contribute to non-ideal interactions. This approximation is suitable for diluted systems for example. However, is not adequate for more general situations where spatial inhomogeneities exist. In contrast, if $\epsilon=0$ the viscous stresses used to solve the macroscale problem are totally computed by the micro-representation, and it implicitly accounts for all stress contributions (ideal and non-ideal) across scales. An important feature of this macro-micro scheme is that allows us to simulate microscopic stresses at arbitrary locations of the macro domain, whereas other regions are modelled using the standard Newtonian discretization. In the results section, we compare the stability and accuracy of \eqref{eq:micromacropi} for different values of $\epsilon$ for different simple and complex fluids. We must remark, that previously reported LL schemes \citep{Murashima2010,Xu2016,Feng2016,Sato2017,Zhao2018, Sato2019,Seryo2020,Morii2020,Schieber2020} correspond to situations where $\epsilon=1$. Hence, assuming that the ideal stress homogeneously contributes over the whole domain from macroscales. Given \eqref{eq:pimean} and \eqref{eq:micromacropi}, we can now express $\bm \tau$ in \eqref{eq:bal} as
%
% \begin{numcases} {\bm \tau =}
% %%  p{\bf I} - \bar{\bm \pi}({\bm v'}, t')  & fully determined microscopic \label{eq:fullmicro}\\
%   p{\bf I} - \Big(\underbrace{\epsilon{\bm \pi^o}({\bm \bar{v}}, \bar{t})}_{\text{macroscopic}} + \underbrace{\Big[(1-\epsilon)\bar{\bm \pi}^o({\bm v'}, t') +\bar{\bm \pi}*({\bm v'}, t')\Big}_{\text{microscopic }}\Big) \label{eq:micromacro}
% \end{numcases}
%
\begin{equation} 
\bm \tau =  -p{\bf I} + \Big(\underbrace{\epsilon{\bm \pi^o}({\bm {v}}, {t})}_{\text{macroscopic}} + \underbrace{\Big[\bar{\bm \pi}^h({\bm v'}, t') - \epsilon \bar{\bm \pi}^o({\bm v'}, t') +\bar{\bm \pi}^*({\bm v'}, t') +\bar{\bm \pi}^k({\bm v'}, t')\Big]}_{\text{microscopic }}\Big). \label{eq:micromacro}
\end{equation}
We must note that in the case where the non-ideal hydrodynamic contributions are negliglible at the microscale, we have from \eqref{eq:pihydro} that $\langle\bar{\bm \pi}^h\rangle=\langle\bar{\bm \pi}^o\rangle$ leading to a simplification of \eqref{eq:micromacro} in the form 
 \begin{equation} 
\bm \tau =  -p{\bf I} +  \Big(\underbrace{\epsilon{\bm \pi^o}({\bm {v}}, {t})}_{\text{macroscopic}} + \underbrace{\Big[(1-\epsilon)\bar{\bm \pi}^h({\bm v'}, t') +\bar{\bm \pi}^*({\bm v'}, t') + \bar{\bm \pi}^k({\bm v'}, t')\Big]}_{\text{microscopic }}\Big) \label{eq:micromacroSimple}
\end{equation}
The principal difference between \eqref{eq:micromacro} and \eqref{eq:micromacroSimple} is that the former requires the microscopic computation of hydrodynamic stresses \textit{at the flow conditions} for both ideal (Newtonian), $\bar{\bm \pi}^o({\bm v'}, t')$, and  complex fluid $\bar{\bm \pi}^h({\bm v'}, t')$. The later, in contrast, only involves the simulation of the hydrodynamic contributions of the investigated fluid. Here, we evaluate our LHMM scheme using \eqref{eq:micromacroSimple}. In section \ref{sec:coupling} we describe the methodology used to estimate the different components of these stresses.  
\\
\\
Considering the representation of fluid in a Lagrangian framework \citep{Espanol2003a}, and the previous decomposition \eqref{eq:micromacroSimple}, the divergence of the total stress in \eqref{eq:bal} takes the form
\begin{equation}
\nabla \cdot {\bm \tau} = -\nabla p + \epsilon \Big(\bar{\eta} \nabla^2 \vcon + \left(\bar{\zeta} + \frac{\bar{\eta}}{D}\right) \nabla \nabla \cdot \vcon \Big) + (1-\epsilon)\nabla \cdot\bar{\bm \pi}^h +\nabla \cdot \bar{\bm \pi}^* +\nabla \cdot \bar{\bm \pi}^k.
\end{equation}
where $D$ is the dimension, and ${\eta}$ and ${\zeta}$ are the shear and bulk viscosities, respectively.
\\
\\
In general, since we aim to incorporate hydrodynamics interactions of the fluid in both scales, a critical requirement for the microscales solver is the capability to model both simple and complex fluids. Here, we model both macro and micro scales using SDPD, discretizing the fluctuating NS equations as a set of $N$ interacting particles with position ${\bf r}_i$ and velocity $\vb_i$.  The system is constituted by particles with a volume $\mathcal{V}_i$, such that ${1}/{\mathcal{V}_i} = d_i = \sum_j W(r_{ij},h)$, being $d_i$ the number density of particles, $r_{ij} = |{\bf r}_i-{\bf r}_j|$, and $W(r_{ij},h)$ an interpolant kernel with finite support $h$ and normalized to one. Additionally, to discretize the NS equations a positive function $\fij$ is introduced such that $\fij = -\nabla W(r_{ij},h)/r_{ij}$. From now, when describing each scale, we identify the discrete particles at microscales with the subindex $i$ and $j$, whereas at macroscale with $I$ and $J$.

\subsection{Macroscales}

At the macroscales, when the volume ${\mathcal{V}}_I$ of the discretizing particle approach continuum scales and thermal fluctuations are negligible, SDPD is equivalent to the smoothed particle hydrodynamics method\citep{Vazquez-Quesada2009}. For this scale, the geometry and type of flow prescribe the boundary condition at $\partial \Omega$. The SDPD discretized equations for \eqref{eq:bal}, describing the particle's position, density, and momentum for a fluid without external forces is expressed as
\begin{numcases}{\rotatebox[origin=c]{90}{\text{Macroscales}}}
 {\der {\bf{r}}_I}/{\der t} = \vcon_I, \label{eq:possph} \\
 {m}\frac{\der {d}_I}{\der t} = \sum_J \fijcon {\rbijcon}\cdot\vcon_{IJ}, \label{eq:rhosph} \\
%\bar{m}\frac{\der \vcon_I}{\der t}  = \sum_J \Big[\bar{\bm \tau}_{IJ} \Big] \fijcon {\rbijcon}
 {m}\frac{\der \vcon_I}{\der t}  = \sum_J \left( \Big[\frac{{p}_I}{d_I^2} + \frac{{p}_J}{d_J^2}\Big] \fijcon {\rbijcon} \right. \nonumber \\
 \hspace{1cm}- \left. \epsilon\Big[\facon \vcon_{IJ} +\fbcon (\vcon_{IJ}\cdot\ebijcon)\ebijcon \Big]  \frac{\fijcon}{d_I d_J} - \bar{\bm \pi}_{IJ} \fijcon {\rbijcon}\right),
 %\Big[(1-\epsilon)(\bar{\bm \pi}_{IJ}) + \bar{\bm \pi}_{IJ}^*\Big] \fijcon {\rbijcon},
%\frac{\bar{\eta}_i+\bar{\eta}_j}{d_i d_j} \vcon_{ij} \fijcon%\frac{\eta_i^2+\eta_j^2}{2\eta_i\eta_j}\fij \rbij,
\label{eq:momsph}
\end{numcases}
where $\rbijcon = {\bm r}_I - {\bm r}_J$, $\vcon_{IJ} = \vcon_I - \vcon_J$, and $\ebijcon= \rbijcon/\rijcon$. In \eqref{eq:momsph}, $\fijcon$ is expressed in terms of the macroscales indicating its correspondence with a interpolation kernel with finite support $\bar{h}$. The term ${p}$ is the density-dependent pressure. $\facon$ and $\fbcon$ are friction coefficients related to the shear ${\eta}$ and bulk ${\zeta}$ viscosities of the fluid through $\facon={(D+2){\eta}}/{D}-{\zeta}$ and $\fbcon = (D+2)({\zeta}+{{\eta}}/{D})$ (for $D=2,3$). 
The microscopically-informed tensor $\bar{\bm \pi}_{IJ}$ is given by
\begin{equation}
\bar{\bm \pi}_{IJ}= (1-\epsilon)\Big(\frac{\bar{\bm \pi}_I^h}{d_I^2} + \frac{\bar{\bm \pi}_J^h}{d_J^2}\Big) + \Big(\frac{\bar{\bm \pi}_I^*}{d_I^2} + \frac{\bar{\bm \pi}_J^*}{d_J^2}\Big) + \Big(\frac{\bar{\bm \pi}_I^k}{d_I^2} + \frac{\bar{\bm \pi}_J^k}{d_J^2}\Big). 
\label{eq:pimacro}
\end{equation}
% \begin{equation}
% \bar{\bm \pi}_{IJ}= \Big(\frac{\bar{\bm \pi}_I^h}{d_I^2} + \frac{\bar{\bm \pi}_J^h}{d_J^2}\Big)-\epsilon\Big(\frac{\bar{\bm \pi}_I^o}{d_I^2} + \frac{\bar{\bm \pi}_J^o}{d_J^2}\Big) + \Big(\frac{\bar{\bm \pi}_I^*}{d_I^2} + \frac{\bar{\bm \pi}_J^*}{d_J^2}\Big) + \Big(\frac{\bar{\bm \pi}_I^k}{d_I^2} + \frac{\bar{\bm \pi}_J^k}{d_J^2}\Big). 
% \label{eq:pimacro}
% \end{equation}
%
The terms $\bar{\bm \pi}_{I}^h$, $\bar{\bm \pi}_{I}^*$, and $\bar{\bm \pi}_{I}^k$ are obtained from the microscale. Their representation is detailed in the subsection \ref{sec:coupling}. 

\subsection{Microscales}

At microscales, the SDPD\citep{Ellero2018} equations contain both deterministic and stochastic contributions. The later accounts consistently for thermal fluctuations. The balance equations are then given by
\begin{numcases}{\rotatebox[origin=c]{90}{\text{Microscales}}}
 {\der {\bf{r'}}_i}/{\der t} = \vb'_i, \label{eq:possdpd} \\
%  m'\frac{\der {d}_i}{\der t} = \sum_j \fijmes \rbijmes \cdot\vmes_{ij}, \label{eq:rhosdpd} \\
m'\frac{\der \vmes_i}{\der t}  = \sum_j \left[ \frac{p'_i}{d_i^2} + \frac{p'_j}{d_j^2}\right] \fijmes \rbijmes
-\sum_j \left[\fa \vmes_{ij} +\fb (\vmes_{ij}\cdot\ebijmes)\ebijmes \right] \frac{\fijmes}{d_i d_j},
\label{eq:deterministic} \\
m'{\der \vbtilde_i} = \sum_j {\left(\Aij \der \wbbarij + \Bij \frac{1}{D}\text{tr}[ \der \wbij] \right) \cdot \ebijmes},
\label{eq:random}
\end{numcases}
where $\vmes_{ij} = \vmes_i - \vmes_j$, $\fa$ and $\fb$ are friction coefficients related to the shear $\eta$ and bulk $\zeta$ viscosities of the fluid through $\fa={(D+2)\eta}/{D}-\zeta$ and $\fb = (D+2)(\zeta+{\eta}/{D})$. Thermal fluctuations are consistently incorporated into the model through the stochastic contributions to the momentum equation by \eqref{eq:random}. Where $\wbij$ is a matrix of independent increments of a Wiener process for each pair $i,j$ of particles, and $\wbbarij$ is its traceless symmetric part, given by
\begin{align*}
\der \wbbarij = \frac{1}{2}\left[d\wbij+\der \wbij^T\right] - \frac{\delta^{\alpha \beta}}{D}\text{tr}[ \der \wbij],
\end{align*}
where the independent increments of the Wiener processes satisfy
\begin{align}
 \der {\bf {W}}_{ii^*}^{\alpha \alpha^*} \der {\bf {W}}_{jj}^{\beta \beta^*} &= [\delta_{ij}\delta_{i^*j^*} + \delta_{ij^*} \delta_{i^*j}]\delta^{\alpha \beta}\delta^{\alpha^* \beta^*} \der t, \nonumber \\
 \der V_{ii^*}\der V_{jj^*} &= [\delta_{ij}\delta_{i^*j^*} - \delta_{ij^*} \delta_{i^*j}]\der t, \nonumber \\ 
\der {\bf W}_{ii^*}^{\alpha \alpha^*} \der V_{ii^*} &= 0.
\end{align}
To satisfy the fluctuation-dissipation balance the amplitude of the thermal noises $\Aij$ and $\Bij$ are related to the friction coefficients $\fa$ and $\fb$ through
\begin{align}
A_{ij} &= \left[4\kbt \fa \frac{\fijmes}{d_i d_j} \right]^{1/2},
\\
B_{ij} &= \left[4\kbt\left(\fb -\fa\frac{D-2}{D}\right)\frac{\fijmes}{d_i d_j} \right]^{1/2},
\end{align}
We remark that in \eqref{eq:random} the prime notation for the $\Aij$ and $\Bij$ is omitted since thermal fluctuations are only accounted for microscales. 
\\
\\
At microscales, \sdpd~ has been used to model complex fluids such as polymer or colloids\citep{Ellero2003, Vazquez-Quesada2009, Moreno2021,Nieto2022} by using additional potentials\citep{Litvinov2008} or constructing colloidal objects with adequate interaction potentials with the surrounding fluid\citep{Vazquez-Quesada2009, Bian2012}. Using this approach, \eqref{eq:deterministic} can be further enlarged, to explicitly account for contributions due to connectivity potentials (i.e. FENE\citep{Litvinov2008}), colloid-solvent interactions\citep{Bian2012}, colloid-colloid interactions\citep{Vazquez-Quesada2009}, blood flow\citep{Moreno2013a, Muller2014, Ye2020}, phase separation \citep{Lei2016}, and coffee extraction\citep{Mo2021}.

\subsection{Coupling}
\label{sec:coupling}

In the proposed LHMM, the transfer of information macro-to-micro occurs through the velocity field of the macroscales, $\vcon$, that defines the boundary conditions of the microscale subsystems. Whereas, the micro-to-macro transfer occurs \textit{via} the stress tensor, $\bar{\bm \pi}$. We denote $\mathcal{N}$ the number of microscopic subsystems generated to compute microscale-informed stresses. In general, $\mathcal{N}$ can be chosen depending on specific macroscopic regions where the stresses need to be computed. However, to facilitate the presentation and validation of the method, we define $\mathcal{N} = \bar{N}$, such that one microscopic simulation is generated per each macroscopic particle. Of course, microscopic simulations contain a large number of microscopic degrees of freedom (e.g. polymers, colloids, droplets) on which the mean average is referred. In general, the total number of degrees of freedom (particles) required to describe a system using LHMM decreases compared to a fully-resolved microscopic system when the length scale separation between scales increases (i.e. towards a continuum representation of the fluid), which offers significant advantages from a computational standpoint. In section \ref{sec:implementation} we further discuss those computational aspects. We present the general stages of the coupling in the figure \ref{fig:coupling} and the Algorithm \ref{alg:lhmm} in the Appendix 1. 

 \begin{figure}%[!tbhp]%tbhp]
	\centering
	\setlength{\unitlength}{0.1\columnwidth}
	\scalebox{0.9}{
	{\includegraphics[trim=0cm 0cm 0cm 0cm, clip=true, width=1\textwidth]{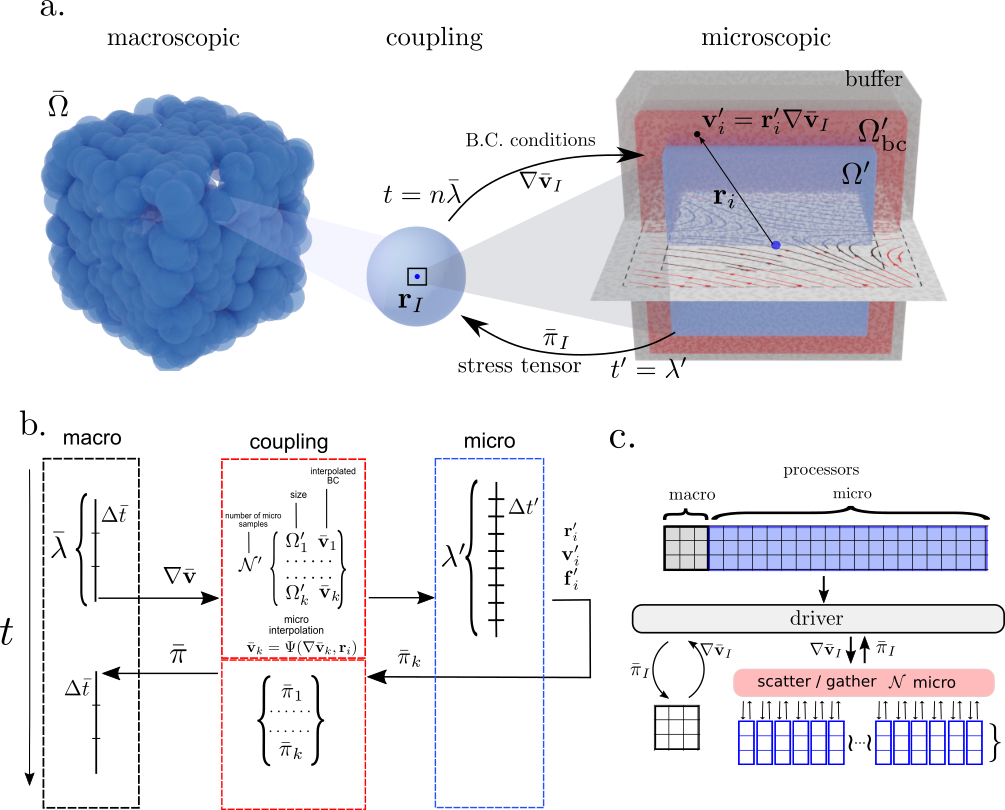}}		
	}	
	\caption{\textit{a.} Schematic representation of the fully Lagrangian heterogeneous multiscale method proposed. \textit{b.} algorithm, and \textit{c.} parallelization}
	\label{fig:coupling}
\end{figure}

\subsubsection{Macro to micro:}
At the microscale, we use a generalized boundary condition scheme recently proposed \citep{Moreno2021} to model arbitrary flow configurations, allowing us to account for non-trivial velocity fields (i.e. mixed shear and extensional). We decompose the microscopic simulation domain in three regions: \textit{buffer}, \textit{boundary-condition} ($\obc$), and \textit{core} ($\omes$), as shown in figure \ref{fig:coupling}. The properties of the fluid, such as the stress tensor, are evaluated from the core region. In the boundary-condition region, the velocity of the particles is prescribed from a macroscopic velocity field $\vcon$. The system is further stabilized and periodic boundary conditions are adopted owing to the buffer region. A detailed description of this domain decomposition approach can be found in \citep{Moreno2021}. To reconstruct the velocity field $\vcon$ at boundary regions $\obc$, we use the velocity gradient $\nabla \vcon_I$ at the macroscale $I$th-particle position. The macroscopic $\nabla \vcon_I$ can be approximated using the SDPD interpolation kernel, such that
% {\color{red} We must note that we adopted this approach due to its flexibility to model arbitrary flow configurations. However, other periodic schemes that use pure shear or extensional flows such as Lees-Edwards\citep{Allen2017} or Kraynik-Reinelt\citep{Kraynik1992} can be also  incorporated in the proposed LHMM framework}
%
\begin{equation}
 \nabla \vcon_I = \sum_J \fijcon \rbijcon \vcon_{IJ}.
 \label{eq:gradv}
\end{equation}
This first-order approximation allows us to compute velocity gradients with a minimal computational cost during the macroscopic force calculation stage. In the results section, we validate the use of this approach. Other high order alternatives to compute $\nabla \vcon_I$, are also possible. However, it would require an additional spatial interpolation step\citep{Zhang2004}. Using \eqref{eq:gradv}, the velocity $\vmes_i$, of the microscale particles located at the boundary-condition region is then determined by
\begin{equation}
 \vcon_i = {\bm r}_i' \nabla \vcon_I, \quad \forall i \in \obc,
\end{equation}
where the macroscopic velocity field is linearly interpolated taking the macroscopic particle centred at the origin of the box (see figure \ref{fig:coupling}). The extent of the microscopic subsystems is given by the characteristic length $\omes$. In general, we consider all microscopic subsystems have the same size $\omes$, however, different sizes can be used, if the specific features of the flow require it.

\subsubsection{Micro to macro:}
Given a macroscopic particle $I$, we determined its stress tensor, $\bar{\bm \pi}_I$, from the microscales. Here, we adopt the  Irving-Kirkwood (IK) methodology\citep{Yang2012} such that the stress is given by $\bar{\bm \pi}_I({\bf x'};t) = \bar{\bm \pi}_I^K ({\bf x'};t) + \bar{\bm \pi}_I^P ({\bf x'};t)$, where $\bar{\bm \pi}_I^K ({\bf x'};t)$ and $\bar{\bm \pi}_I^P ({\bf x'};t)$ account for kinetic and potential contributions to the stress tensor, respectively. This potential contribution contains both hydrodynamic and non-hydrodynamic terms. We use the weighting function $w_{IK}({\bf r'},{\bf x'})$ for the spatial averaging, whereas time averaging is conducted over a range on  $N_t'$ microscopic time steps. The number of time steps used to perform the averaging typically spans the duration of the microscale simulation.  The kinetic part is then given by \citep{Tadmor2011} 
\begin{equation}
\pima_I^K ({\bf x};t) = -\frac{1}{N_t'}\sum\limits_{n=1}^{N_t'} \left[ \sum\limits_{i} m_i \, w_{IK}({\bf r}_i(n)-{\bf x}){\vartriangle \vmes_i(n)}\otimes {\vartriangle \vmes_i(n)} \right],
\label{eq:skin}
\end{equation}
where $\vartriangle \vmes_i(n) = \vmes_i - \left\langle \vmes ({\bf r}_i;n) \right\rangle $ is the relative velocity of the particle $i$ at time step $n$. In the IK approach, if $\varphi_{ij}$ is the magnitude of the force between particles $i$ and $j$, it is considered that the force term can be expressed in central force decomposition as
\begin{align*}
{\bf f}_{ij}(n) \otimes \rbij(n) = \frac{\varphi_{ij}(n)\rbij(n)}{\rij(n)},
\end{align*}
With this, the potential part of the stress tensor reads
\begin{align}
\pima_I^P ({\bf x},t) = \frac{1}{2N_t'}\sum\limits_{n=1}^{N_t'} \left[\sum\limits_{\substack{i,j \\ i\neq j}} {\bf f}_{ij}(n) \otimes \rbij(n) \, \mathcal{B}({\bf x};{\bf r}_i(n),{\bf r}_j(n)) \right] 
\label{eq:spot}
\end{align} 
where $\mathcal{B}({\bf x};{\bf r}_i(n),{\bf r}_j(n))$ is a bond function given by $\mathcal{B}({\bf x; u,v}) = \int_{s=0}^1 w_{IK} \left((1-s){\bf u}+s	{\bf v}-\bf{x} \right) \der s$. The bond function is the integrated weight of the bond for a weighting function centred at $\bf x$. If the weighting function $w_{IK}({\bf y}-{\bf x})$ is taken as constant within a domain $\Omega_a$, and zero elsewhere, $w_{IK} = 1/\text{Vol}(\Omega_a)$ if ${\bf y} \in \Omega_a$. If additionally, the bond function $\mathcal{B}$ is calculated only with bonds fully contained in $\Omega_a$, we would have $\mathcal{B}({\bf x};{\bf r}_i,{\bf r}_j) = 1/\text{Vol}(\Omega_a)$ for $i-j \in \Omega_a$. For more detailed descriptions and extended validation benchmarks, we refer the reader to \cite{Moreno2021}.

% \begin{multline}
% \pima_I^o ({\bf x},t) = 2\eta \frac{1}{N_t'}\sum\limits_{n=1}^{N_t'} \sum\limits_{\substack{i,j \\ i\neq j}} \frac{1}{2}\Big[(\langle \vmes_{j}\rangle-\langle \vmes_{i}\rangle) \otimes \textbf{e}_{ij} m_{j} w_{IK}({\bf r}_i(n)-{\bf x})  \\
% +  \Big( (\langle \vmes_{j}\rangle-\langle \vmes_{i}\rangle ) \otimes \textbf{e}_{ij}\Big)^{T} w_{IK}({\bf r}_i(n)-{\bf x}) \Big]
% \end{multline}

% The summary of the proposed coupling methodology is presented in figure \ref{fig:coupling}

\subsubsection{Time-stepping}

\begin{figure}%tbhp]
	\centering
	\setlength{\unitlength}{0.1\columnwidth}
	\scalebox{0.9}{
	{\includegraphics[trim=0cm 0cm 0cm 0cm, clip=true, width=1\textwidth]{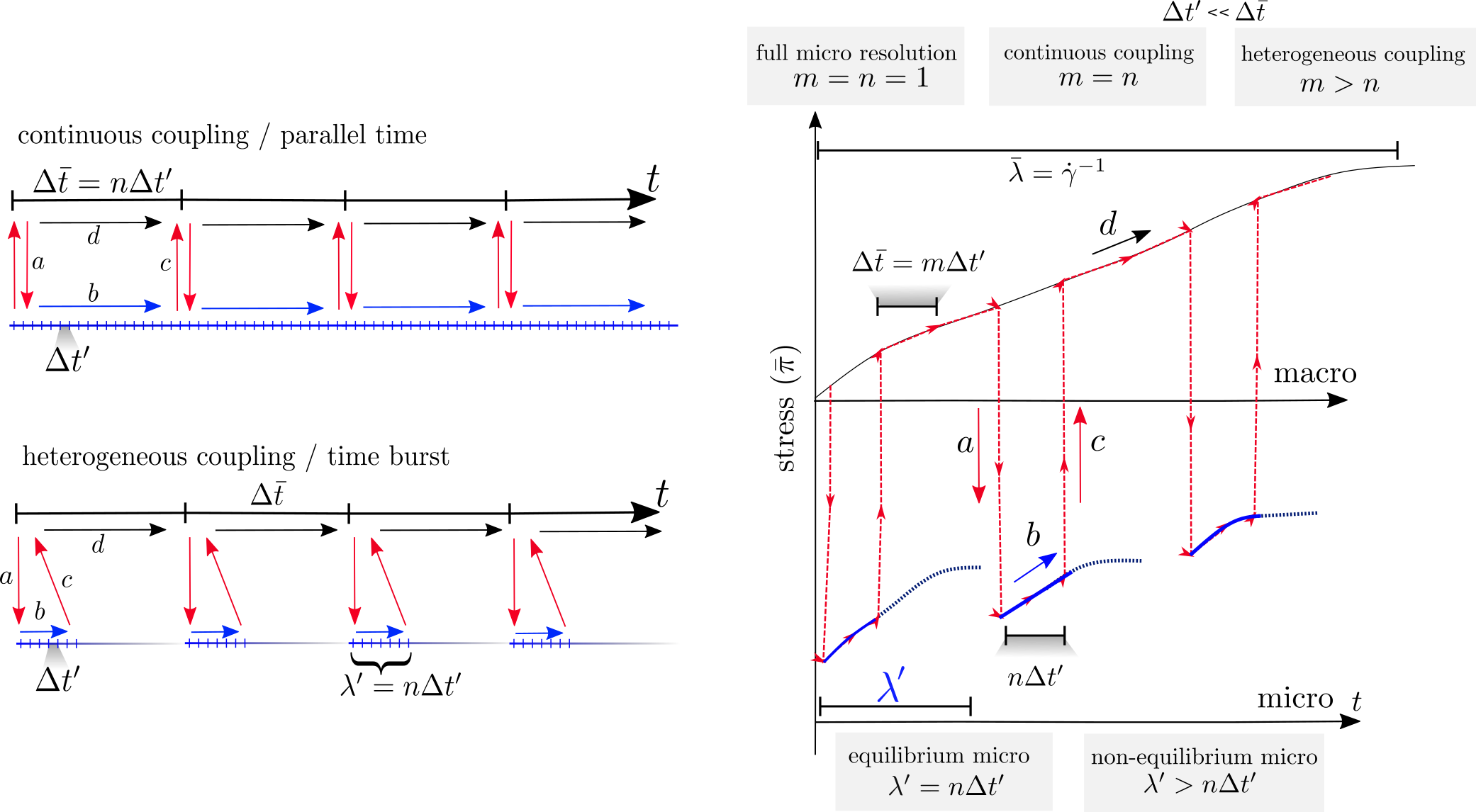}}		
	}	
	\caption{Time-stepping approaches and information passing between scales. In LL-schemes is in principle is possible to pass information from micro solvers before reaching full equilibration, since the historically-dependent stress is naturally tracked in the Lagrangian framework.}
	\label{fig:timestep}
\end{figure}

A critical aspect of heterogeneous multiscale methods is the time-stepping approach used to send information between scales\citep{E2009, Lockerby2013}. From macroscales, we consider the time step is given by $\dtmac$, whereas the overall time scale $\lambda_{MM}$ of the system investigated is related to the operative conditions, such as the shear rate, $\dot{\gamma}$. Thus macroscopic scales define the extent of the overall simulations, requiring a minimum of $m$ steps ($\lambda_{MM} = m \dtmac$). The time-stepping approach depends on the time-scale separation between macro and microsystems. If we denote the characteristic relaxation time for each scale as $\lambda$, systems with large time-scale separation satisfy $\lammic \ll \lammac$, whereas for highly coupled scales $\lammic \approx \lammac$. From microscales, the time step $\dtmic$, sets the condition to accurately resolve the stress evolution of the system. The relaxation of the microscales requires a minimal number of timesteps $n$, such that $\lammic = n  \dtmic$.  In practice, microscopic simulations would use $n$ large enough ($\lammic < n  \dtmic$) to ensure the proper stabilization of the system and to reduce the noise-to-signal ratio. 
\\
\\
In multiscale methods, the relaxation time of the macro and micro systems determines the ratio $\dtmac/ \dtmic$. As the limit condition for the highest temporal resolution we can consider the case of $\dtmac/ \dtmic = 1$. However, in practice, this would not correspond to a temporal multiscale method, but a fully microscopic description of the system. In those cases, the gain in performance for using HMM comes only from the spatial upscaling of the stress. Existent LL schemes \citep{Yasuda2014,Sato2017, Sato2019} that use time steps in the same order for macro and micro solvers are limited to problems with microscale temporal resolutions. Otherwise, in the case of stochastic microscale simulations\citep{Morii2020}, equilibration of the microscales is assumed through mean-field approximations. Due to these practical restrictions, different time-stepping approaches have been recently investigated \citep{E2009,Lockerby2013} to increase the temporal gain in HMMs and reach macroscopic time scales.  Depending on the order of magnitude of $\dtmac/ \dtmic$, different time-stepping schemes can be used. In figure \ref{fig:timestep}, we illustrate the basic sequence of time stepping: $a)$ scattering $\nabla \vcon_I$ on individual microscopic solvers; $b)$ solving microscales under arbitrary BC; $c)$ gathering $\bar{\bm \pi}_I$ for macroscales; and $d)$ solving macroscales. The simplest time-stepping, typically referred as \textit{continuous coupling} between scales (see figure \ref{fig:timestep}), considers that micro solvers are evolved during $n\dtmic$, whereas the time integration at macroscale occurs at $\dtmac = n\dtmic$. An alternative to achieve both spatial and temporal gain when using our LL schemes is the heterogeneous-coupling time stepping \citep{Lockerby2013} (a.k.a time burst), as presented in figure \ref{fig:timestep}. In time-burst approaches, the macroscales are evolved using $\dtmac = m \dtmic$, where $m>>n$. Therefore, microscale behaviour is extrapolated over larger periods. Compared to continuous coupling, the overall gain of heterogeneous time stepping is given by the ratio $m/n$. In general, for highly coupled scales ($\lammic \approx \lammac$) we would require $m \sim n$, to reach the continuous coupling. EE and EL schemes with time-burst time stepping have been adopted for systems with large enough time-scale separation ($\lammic \ll \lammac$). However, due to the incompatibility of simple Eulerian description to capture memory effects, this approximation of constant microscopic stresses over a larger macroscopic time exhibit larger deviations as the microscale relaxation time increases. These limitations can be significantly relieved using LL-schemes\citep{E2009}. Here, depending on the type of system and scale separation, we used both continuous and heterogeneous coupling in time. 
\\
\\
% Indeed, in a Lagrangian framework, is possible to pass information from micro solvers before reaching full equilibration, since the complete flow history is naturally tracked for each macroscopic element.
The Lagrangian nature of the proposed framework represents a critical ingredient to perform the multiscale coupling with SDPD. Flow history is by default accessible to every element of fluid (SPH particle), which carries its microstructure (in a Lagrangian sense). As a consequence, the initial conditions (SDPD positions/velocities) at every macroscopic time step can be taken as those at the end of the previous time step, regardless of whether the microstructure has relaxed or not within it. This idea allows us to apply HMM directly to the flow of complex fluids by running SDPD simulations in parallel (one for each SPH particle) undergoing inhomogeneous and possibly unsteady velocity gradients obtained from the macroscopic SPH calculation. As discussed in [Bertevas et al. (2009)], accurate IK estimates in mesoscopic calculations require typically periodic representative elementary volumes (RVE) three to ten times larger in linear size than the suspended solid particles, and therefore we expect a significant computational gain when applying this procedure to SPH fluid volumes much larger than the RVE.  

% For highy couple systems, where  $\lammic \approx \lammac$, 
%  Here, we focus on multiscale schemes that used $\dtmac/ \dtmic >1$.
% In general, $\lammic \approx \lammac$ involves the evolution of the microstructures on temporal scales comparable with macroscales, thus poses practical challenges for HMM. 
%  Continuous coupling are computationally restrictive, and . 

% The use of a fully Lagrangian approach bypass some of the limitations associated with Eulerian frameworks\citep{Enquist} as each discretized point of fluid will transport explicitly their microscopic stress. 

% To streamline the information-passing between scales we introduce a dimensionless stress given by 
% $\frac{ \bar{bm \pi'} \dot{\gamma}'}{\eta'}$. 

%  \begin{figure}%[!tbhp]%tbhp]
% 	\centering
% 	\setlength{\unitlength}{0.1\columnwidth}
% 	\scalebox{0.8}{
% 	{\includegraphics[trim=0cm 0cm 0cm 0cm, clip=true, width=1\textwidth]{./timeSeparation.png}}		
% 	}	
% 	\caption{Information passing scheme for two different microscopic relaxation times. For large enough time-scale separation time burst approach is suitable, due to the fast relaxation time at the microscales. In contrast for systems with larger relaxation times, the seamless methodology is adopted. For the later, macro and micro run co-currently exchanging infromation every time step. However, each scale runs asynchronously (different time scales).}
% 	\label{fig:timeseparation}
% \end{figure}

\subsection{LHMM Implementation}
\label{sec:implementation}

Since each macroscopic particle is equipped with its microscale solver, the overall cost of the HMM simulations increases compared to constitutive-equations-based approaches. However, the expected cost is significantly reduced for fluids that require to be solved with a resolution at the microscopic scale (polymer coil or colloid scale for example). LL schemes offer parallelization advantages, allowing each macro particle to compute its stress independently. Here, we implement the LHMM using a \textit{c++} driver, coupled with multiple parallel instances of LAMMPS\citep{Plimpton1995} to solve both macro and micro scales. In figure \ref{fig:coupling}$.c$ we illustrate the parallelization approach used. An important feature of the current implementation is that both macro and micro scales can be fully parallelized separately. This has significant advantages compared to fully microscopically resolved systems. In those, the computational cost does not scale linearly as the size of the macroscopic domain reaches continuum scales.        
\\
\\
Since both scales are solved using SDPD, we can estimate the relative cost of solving a given system in terms of the total number of discretizing particles used or degrees of freedom (DOFs). Considering a macroscopic system of size $\bar{L}$ being fully microscopically resolved with interparticle distance $\dhmi$, then the total number of DOFs is given by $N_{\text{full}} = (\bar{L}/\dhmi)^D$, where $D$ is the dimension of the system. This system in a LHMM discretization requires $N_{\text{LHMM}} = \bar{N}N'$ total particles, where $\bar{N} = (\bar{L}/\dhma)^D$ and $N' = (\Omega'/\dhmi)^D$, being $\Omega'$ the size of the microscopic domain sampled. Additionally, if we define the spatial and temporal gain of the LHMM method as $G_s = \bar{h}/\Omega'$ and $G_t = \dtmac/\dtmic$, respectively. The total number of DOFs for LHMM can be expressed as
\begin{equation}
 N_{\text{LHMM}} = N_{\text{full}} \Big(\frac{\Omega'}{\dhma}\Big)^D = N_{\text{full}} \frac{\kappa}{G_s},
\label{eq:Nlhmm}
\end{equation}
where the ratio ${\Omega'}/{\dhma}$ is inversely proportional to the spatial gain $G_s$ achieved by the LHMM, since at the macroscale $\bar{h} = \kappa \dhma$. The value of $\kappa$ is typically determined by the required number of neighbour points for the kernel interpolation and is related to the accuracy of the method\citep{Ellero2011}. Herein, we use $\kappa=4$ \citep{Bian2012} (for both macro and micro scales). From \eqref{eq:Nlhmm}, we can readily identify that compared to a fully resolved system the LHMM entails a reduction in DOFs required for systems with $G_s>4$. In general, the goal of HMM is to model systems with spatial gains orders of magnitude larger to tackle continuum scale problems with microscopic detailed effects.
\\
\\
Another computational gain associated with the LHMM is the flexibility of using larger time steps compared to a fully-resolved system. The Courant-Friedrichs-Lewy (CFL) condition determines the stability criterium for the minimum integration time step for microscales, $\dtmic = \dhmi/c$, where $c$ is the artificial speed of sound. As discussed in the previous section, for a target macroscopic time scale $\lambda_{MM}$, the total number of times steps required is then $n_{\text{full}} = \lambda_{MM}/\dtmic = (c\lambda_{MM})/\dhmi$. Thus, for instance, to model a system on the order of seconds with a nanoscopic resolution would typically require $n_{\text{full}} \propto 10^{12}$ time steps. In LHMM, the CFL condition at the macroscale allows the use of $\dtmac = \dhma/c \propto G_s\dtmic$, that scales with the spatial gain, it is in principle feasible to integrate macroscale equations over significantly larger time steps. It is worth noting, that a slightly smaller macroscopic time steps may be preferred to comply with the characteristic microscopic relaxation time, as discussed in the previous section. In HMM, the temporal gain is in general limited by the capability of the method to accurately keep track of the historically dependent stress. This aspect is an important feature of the proposed fully Lagrangian scheme, allowing the use of larger macroscopic time steps, compared to Eulerian-Lagrangian settings.    
% Henceforth, $n_{\text{macro}} = (c\lambda_{MM})/ 

% 
%  \begin{figure}%[!tbhp]%tbhp]
% 	\centering
% 	\setlength{\unitlength}{0.1\columnwidth}
% 	\scalebox{0.5}{
% 	{\includegraphics[trim=0cm 0cm 0cm 0cm, clip=true, width=1\textwidth]{./parallelization.png}}		
% 	}	
% 	\caption{Scheme of parallelization}
% 	\label{fig:parallelScheme}
% \end{figure}

\section{Macro and micro descriptions}

\begin{figure}%tbhp]
 	\centering
 	\setlength{\unitlength}{0.1\columnwidth}
 	\scalebox{0.95}{
 	{\includegraphics[trim=0cm 0cm 0cm 0cm, clip=true, width=1\textwidth]{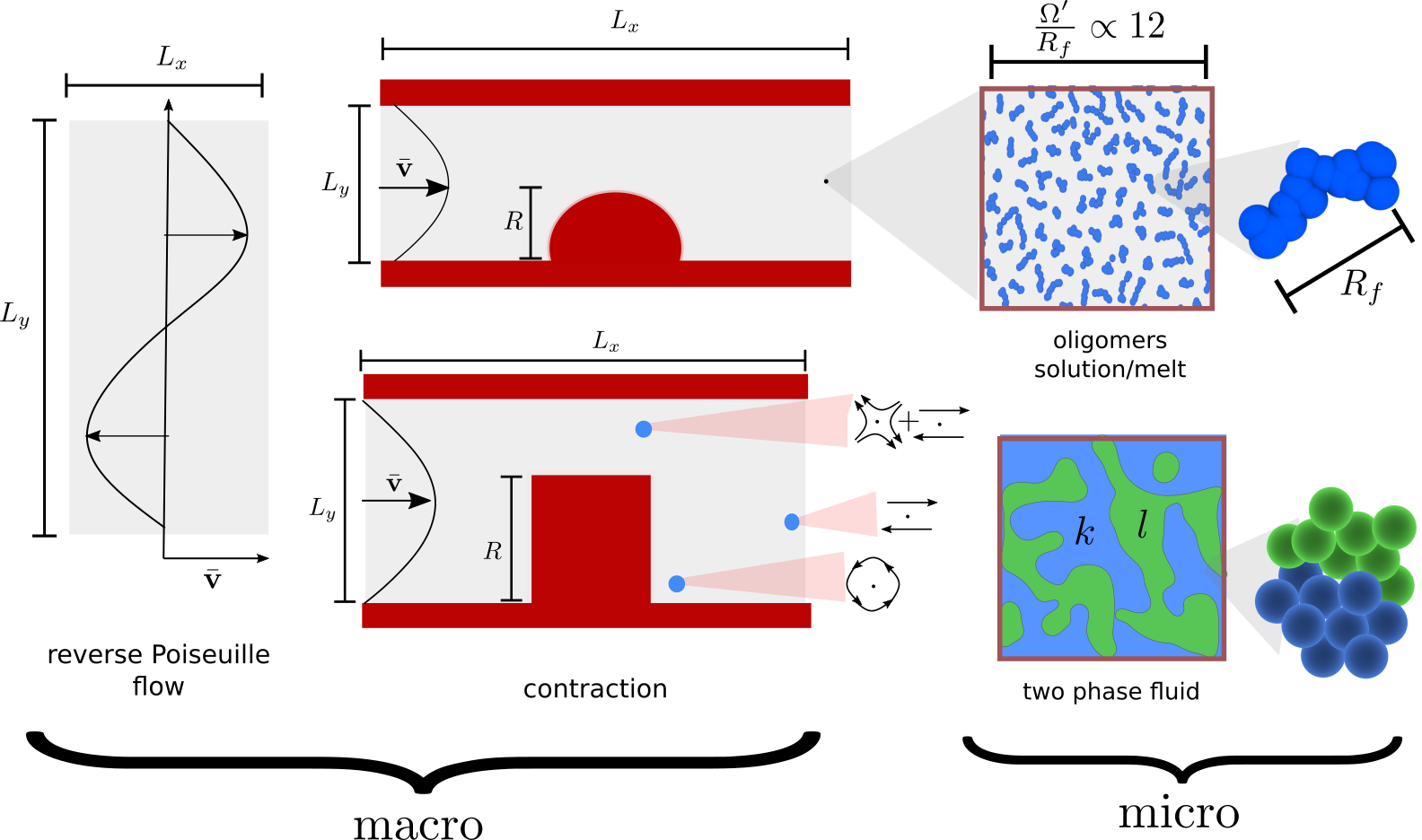}}		
 	}	
 	\caption{Sketch of macro and microscopic systems investigated. At the macro scale reverse Poiseuille flow and cavity flows (cylindrical and square) are constructed. For micro scales, in addition to the standard Newtonian fluid, complex fluids are modelled as oligomeric solutions and melts, and two immiscible fluids $k$ and $l$, undergoing microphase separation.}
 \label{fig:complexG}
 \end{figure}

We conduct a series of different benchmark tests for a simple Newtonian fluid to validate the consistency and stability of the proposed multiscale method. We consider a macroscale system under reverse Poiseuille flow \citep{Fedosov2010b} in a domain of size $L_y \times L_x$, and evaluate the effect of the stabilizing parameter $\epsilon$, on the range $[0-1]$. Additionally, we evaluate the proposed LHMM framework on other geometries that induce different local flow types (i.e. shear, extension, and mixed flow)  corresponding to a flow in circular and square contraction arrays. The use of arbitrary BC \citep{Moreno2021} at the microscales allows us to account for different spatial flow configurations. In figure \ref{fig:complexG} we summarise the type of flow configurations investigated. For circular contraction arrays, the size of the channel was $L_y=16\bar{h}$ and $L_x=20\bar{h}$, whereas the size of the contraction is $R=4\bar{h}$. At the walls, we adopt the methodology used by \cite{Bian2012}, such that the velocity of the wall particles used to compute the viscous forces is extrapolated to enforce non-slip boundary conditions, $\vcon=0$ at the fluid-wall interface. 

\com{To illustrate the flexibility of the proposed LHMM framework at the microscales we model various physical problems. We adopt different generic SDPD models for polymeric and multiphase systems. We must note that these complex fluids are used here only to showcase our multiscale methodology, thus, a systematic parametric analysis of the specific systems is out of the scope of this work, and will be addressed in future publications.}

\subsection*{Oligomer melts and solutions}
We model non-Newtonian fluids by constructing melts and solutions of oligomers of $N_s=8$ and $N_s=16$ connected SDPD particles. We use finitely extensible nonlinear elastic (FENE) potential of the form  $U_{\text{fene}} = -1/2 k_s r_s^2 \text{ln} \left[1-(r/r_s)^2\right]$, where $k_s$ and $r_s$ are the bond energy constant and maximum distance, respectively. In our simulations we fix $k_s = 23k_BT/r_s^2$ and $r_s=1.5\dhmi$. We characterize the oligomers in the system through its end-to-end vector $\textbf{R}_f$, to determine the mean end-to-end distance $\langle R_f \rangle^2 = \langle |\textbf{R}_f|\rangle^2$. The measured equilibrium end-to-end radius, $R_f$, under no flow condition is $R_f = 0.3\pm 0.02$. Given the size of the microdomain and oligomers, the microscales are being sampled on size ratios $10 < \omes/R_f \sim <13$ approximately. Polymeric systems constructed in simular fashion in SDPD\cite{Nieto2022} have shown that the polymer relaxation times $\lambda_p$ agreed with the Zimm model. Herein, we identify relaxation times for $N_s = 8$ on the order of $\lambda_p \approx 6 t_{\text{SDPD}}$, and for $N_s=16$ on the order of $\lambda_p \approx 9 t_{\text{SDPD}}$. The Weissenberg numbers ($Wi=\dot{\gamma}\lambda_p$) investigated on the different examples, full micro and multiscale, ranged from $0.3$ to $100$.     

\subsection*{Two phase flow}

We also constructed microscale systems constituted by two immiscible phases $l$ and $k$. The composition of each phase is denoted, $\kappa_p$, for $p=k,l$, such that the binary mixture satisfies, $\kappa_l + \kappa_k =1$. We adopt the SDPD scheme proposed by \cite{Lei2016} for multiphase flows. In this scheme the momentum equation at microscale \eqref{eq:deterministic} incorporates an additional pairwise term $F^{\text{int}}$, that account for interfacial forces between two phases $k$ and $l$, such that
\begin{align}
 F_{ij}^{\text{int}} = -s_{ij} \phi(\rij) \frac{\rbij}{\rij}
 \label{eq:multiphase}
\end{align}
where 
\begin{equation}
s_{ij} = 
\begin{cases}
s_{kl}, & {\bf r}_i \in \Omega_k \quad \text{and} \quad {\bf r}_j \in \Omega_l, \\
s_{kk}, & {\bf r}_i \in \Omega_k \quad \text{and} \quad {\bf r}_j \in \Omega_k, \\
s_{ll}, & {\bf r}_i \in \Omega_l \quad \text{and} \quad {\bf r}_j \in \Omega_l, \\
\end{cases}
  \label{eq:surfaceT}
\end{equation}
and 
$\phi(\rij)$ is a shape factor given by
\begin{equation}
 \phi = \rij \Big[-Ge^{-\frac{\rij^2}{2r_a^2}} + e^{-\frac{\rij^2}{2r_b^2}} \Big],
\end{equation}
where $G=2^{D+1}$, being $D$ the dimension. The range for repulsive and attractive interactions are defined as $2r_a = r_b = \rho_n^{-1/D}$, such that a relative uniform particle distribution are obtained for a given interfacial tension $\sigma$. The interaction parameters satisfy $s_{kk} = s_{ll} = 10^3s_{kl}$, and the magnitude can be obtained from the surface tension and particle density of the system as
\begin{equation}
s_{qq} = \frac{1}{2(1-10^{-3})}\rho_n^{-2} \frac{\sigma}{[4-D]^{-1}([4-D]\pi)^{1/[4-D]} (-Gr_a^{D+3}+r_b^{D+3})}.
\end{equation}
Here, we model the multiphase systems considering a viscosity ratio between both phases $\eta_k/\eta_l = 1$, and interfacial tension $\sigma = 0.5$. The characteristic time, $\lambda_{ps}$, for total phase separation of a phase $k$ with concentrations of $0.2$ and $0.5$ (starting from a homogeneous mixture), were identified as $\sim 140 t_{\text{SDPD}}$ and $\sim 40 t_{\text{SDPD}}$, respectively. In general, the size ($4h < \Omega' <10h$) of microscale systems investigated and shear rates used, leads to capillary numbers $Ca = (\eta \dot{\gamma} \Omega')/(2\sigma)>10$, that are typical for highly deformable and breakable droplets of the suspending phase \citep{Kapiamba2022}. {\color{black}
It has been shown experimentally that at low $Ca$ numbers, the steady state morphology of multiphase systems can be described as a single value function of the flow. However, when microstructural properties are determined by the balance between break-up and coalenscence of the phases, the morphology can be controlled by the initial conditions of the system, leading to more than one steady state morphology \citep{Minale1997}.}

\section{Results and discussion}

The proper estimation of the velocity gradient at macroscales as well as the correct measurement of the stress tensor from microsystems are key components of the proposed LHMM. Therefore, before validating a fully coupled LHM system, we verify that numerical errors associated with particle resolution at each scale are negligible and that the arbitrary boundary conditions used for microscales do not introduce spurious artifacts on the stress for complex systems.  

\subsection{Macroscopic particle resolution, velocity gradient and stress tensor interpolation}

\begin{figure}%[!h]%tbhp]
	\centering
	\setlength{\unitlength}{0.1\columnwidth}
	\scalebox{0.6}{
	{\includegraphics[trim=0cm 0cm 0cm 0cm, clip=true, width=1\textwidth]{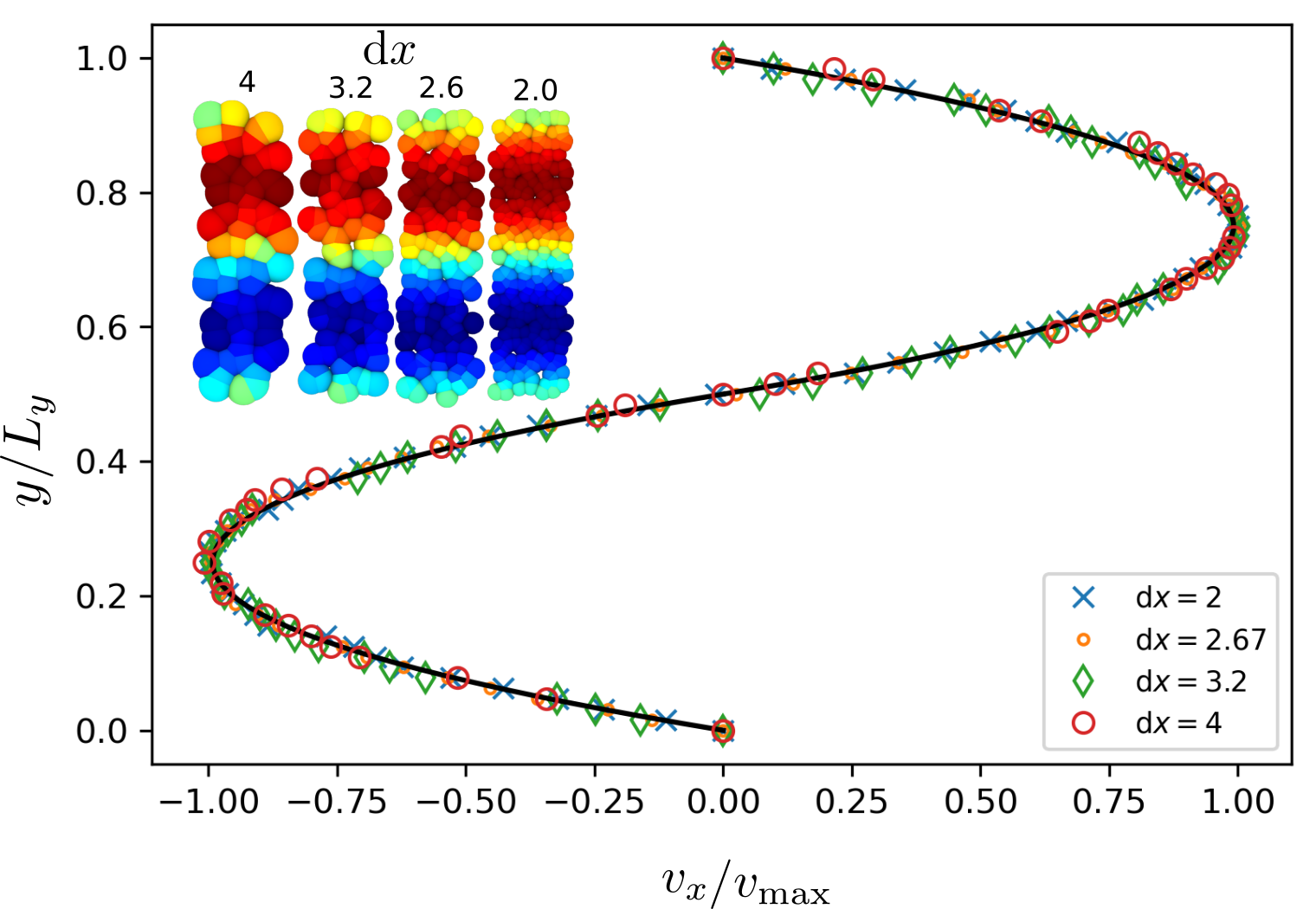}}		
	}	
	\caption{Velocity field for RPF configurations for four different macroscopic particle resolution, corresponding to a total number of particles ${N}=[64,100,144,256]$}
	\label{fig:macroV}
\end{figure}

We determine the minimal macroscopic resolution required to capture the characteristic velocity profile in a reverse Poiseuille flow. We validate the convergence of the velocity field in a domain of size $0.25L_y \times L_y$ with $L_y=64$, for different particle resolutions $L_y/\dhma = [16, 20, 24, 32]$. The obtained velocity profiles are presented in figure \ref{fig:macroV}. From these tests, we identify that even at lower resolutions, $L_y/\dhma = 16$, the accuracy of the profile is acceptable for practical purposes. Hereinafter, we evaluate the proposed LHMM using macroscopic resolutions $L_y/\dhma =16$ and $20$, as a good compromise between minimal numerical error and lower computational cost.   

\begin{figure}%[!h]%tbhp]
	\centering
	\setlength{\unitlength}{0.1\columnwidth}
	\scalebox{0.8}{
	{\includegraphics[trim=0cm 0cm 0cm 0cm, clip=true, width=1\textwidth]{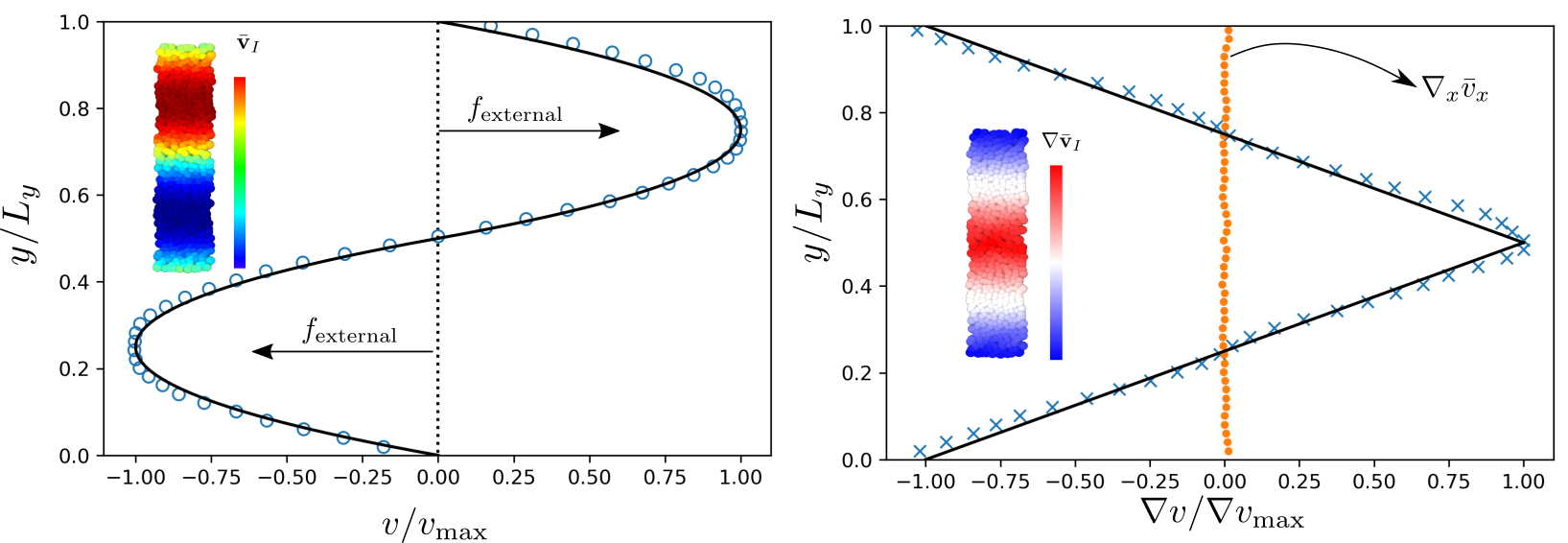}}		
	}	
	\caption{Imposed velocity field and the corresponding gradient for macroscales (solid line), compared with the computed values for each particle $I$ in a domain $10\dhma \times 50\dhma$. This case corresponded to a full macroscale Newtonian fluid with $\epsilon = 1$.}
	\label{fig:vgradv}
\end{figure}

As discussed in the coupling section, we used \eqref{eq:gradv} to compute the macroscopic velocity gradient. We verified this approximation to $\nabla \vcon$ in a RPF, for a macroscopic domain of size $10\dhma \times 50\dhma$. In figure \ref{fig:vgradv}, we present the velocity and components of the velocity gradients (i.e. $\nabla_y v_{x}$ and $\nabla_x v_{x}$) measured, along with the theoretical solutions. Overall, we identified that \eqref{eq:gradv} provides up to a good approximation of the macroscopic velocity gradient required to define the boundary conditions of the microscale simulations. Even though, more refined alternatives to compute $\nabla \vcon$ exist\citep{Zhang2004}, such refinements are out of the scope of the present work.

% The divergence of the microscopically-informed tensor on macro scales is achieved using \eqref{eq:pimacro}.

\begin{figure}%[!tbhp]%tbhp]
	\centering
	\setlength{\unitlength}{0.1\columnwidth}
	\scalebox{1.0}{
	{\includegraphics[trim=0cm 0cm 0cm 0cm, clip=true, width=1\textwidth]{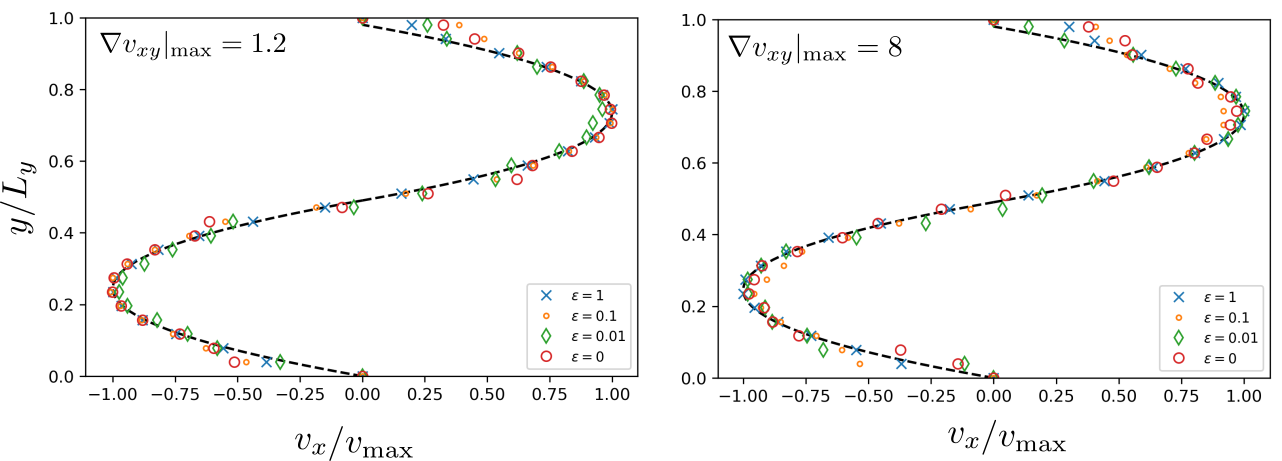}}		
	}	
	\caption{Velocity field for macroscales for different values of $\epsilon$, using the manufactured-microscopic solution of a Newtonian fluid, from the analytical solution for the stress tensor, $\bm \pi^h = \eta (\nabla \vcon + \nabla^T \vcon) + (\zeta - 2\eta/D)\nabla \cdot \vcon \textbf{I}$. The dashed line indicates the theoretical parabolic profile. The results correspond to two systems with different maximum velocity gradient, $\nabla_y v_{x}|_{\text{max}}$.}
	\label{fig:vgradvEps}
\end{figure}

At the macroscales, the divergence of the stress tensor $\nabla \cdot {\bm \tau}$ considers the SDPD interpolation of the microscopically-informed tensor, $\bar {\bm \pi}_{IJ}$, and the stabilizing parameter, $\epsilon$, according to \eqref{eq:momsph}. The accuracy of such interpolation without the numerical errors associated with the actual microscales subsystems is estimated using the analytical solution of a Newtonian fluid. This allows us to manufacture microscopic solutions of $\bm \pi$ to solve micro-macro simulations. The analytical solution of the stress tensor is given by
\begin{equation}
\bm \pi^h = \eta (\nabla \vcon + \nabla^T \vcon) + (\zeta - 2\eta/D)\nabla \cdot \vcon \textbf{I}. 
\end{equation}
Thus, we can compute $\nabla \cdot {\bm \pi}^h$ using the velocity gradients determined on macroscales. In figure \ref{fig:vgradvEps}, we present the velocity profile for systems with various values of $\epsilon$, for  different $Re$, corresponding to magnitudes the maximum velocity gradient $\nabla_y v_{x}|_{\text{max}} = {1.2}$ and $8$. At the evaluated Reynolds numbers and velocity gradients, the flow can be adequately modelled using only the manufactured microscopic solutions ($\epsilon \approx 0$), this is, macroscopic stress tensor can be recovered from microscale systems, with minimal interpolation errors at the macroscale. In general, we observe that at modest values of $\epsilon$ it is possible to fully recover the behaviour of the fluid. 
% It is worth noting that the stress contributions from microscales are crucial to recovering the proper velocity field. This is shown in figure \ref{fig:vgradvEps} for $\nabla_y v_{x}|_{\text{max}} = 1.2$ with $\bar{\bm \pi}_{IJ}=0$ and $\epsilon=0.1$. In this case, the macro contribution only accounts for ten per cent of the viscous stress, whereas the microscopic counterpart is set at zero to highlight its relevance.          

\subsection{Microscales under rigid rotations}

\begin{figure}%[!h]%tbhp]
	\centering
	\setlength{\unitlength}{0.1\columnwidth}
	\scalebox{1}{
	{\includegraphics[trim=0cm 0cm 0cm 0cm, clip=true, width=1\textwidth]{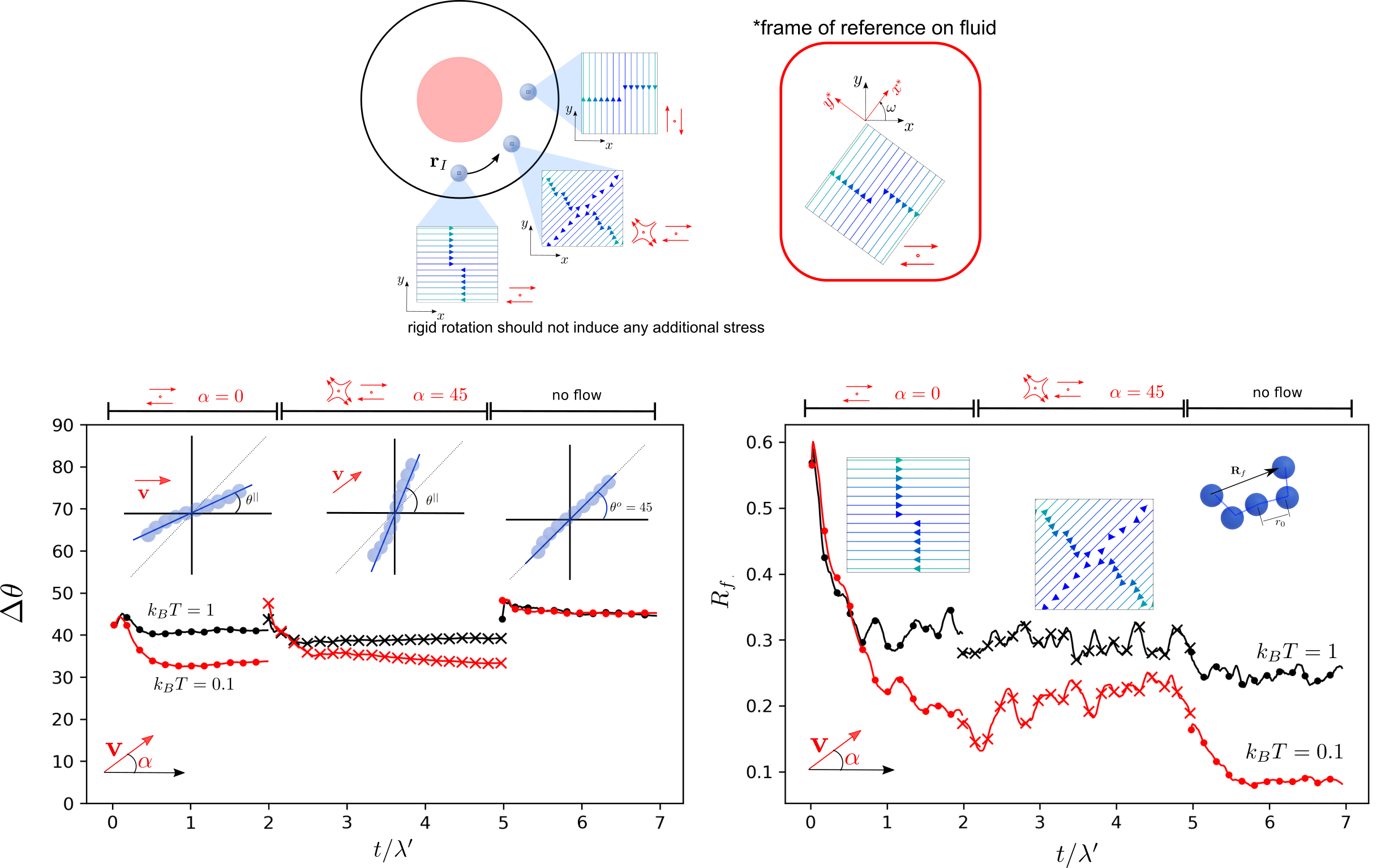}}		
	}	
	\caption{Effect of rigid rotation for an oligomeric microscale system using arbitrary boundary condition scheme. \textit{a)} Schematic of a macroscopic particle undergoing rigid rotation, and the corresponding applied velocity gradient as the particle moves. For comparison, we include the corresponding velocity field when the reference frame is aligned with the particle velocity. \textit{b)} Variation of the mean orientation angle and \textit{c)} mean end-to-end distance $R_f$ of the oligomer coils of size $N_s=8$, in a simulation domain that is rotating from pure shear to $\alpha=\pi/4$ and finally under no flow. Here $\lammic$ denotes the relaxation time of the system.}
	\label{fig:rotation}
\end{figure}

As presented by \cite{Moreno2021}, complex flow patterns can be easily implemented at the microscales to determine the stresses. In LHMM each microscale system may experience temporal variations of the applied velocity gradient even under steady flow conditions, as they travel within an inhomogeneous macroscopic domain, along Lagrangian trajectories. The velocity gradients imposed on microscopic simulations are then referred to a fixed reference frame in the macro domain (see figure \ref{fig:rotation}.\textit{a}). The use of this reference frame leads to microscopic systems that experience transitions from simple shear to mixed shear-extension as the macroscopic particle rigidly rotates. This transition of course should originate from an affine rotation on the measured stress. However, it should not generate any change in the state of stress of the system. As a consequence, an important attribute to verify from the boundary condition scheme of \cite{Moreno2021} is that rigid rotations on the velocity field applied on the boundary condition domain do not alter the microstructure and rheological properties of the fluid.
\\
\\
As a validation test, we construct microscale simulations for oligomeric systems and determine the response of the system as the applied field experiences a large rigid rotation of $45^o$ ($\alpha=\pi/4$). We consider a generalized velocity field over the boundary region of the form
\begin{align}
 \vcon = 
\begin{pmatrix}
\edot x  & \gdot y & 0 \\
\gdoty x & -\frac{1}{2}\edot(1+q) y & 0 \\
0 & 0 & -\frac{1}{2}\edot(1-q) z
\end{pmatrix},
\label{eq:velTensor}
\end{align}
where $q$ is a free parameter, $\edot$ and  $\gdot$ are the strain and shear rate, respectively. The values of $\edot$, $\gdot$, $\gdoty$, and $q$ define the flow configuration\citep{Bird1987}. The velocity gradient rotated by an angle $\alpha$ is given by $\nabla\bar{\bf{v}}(\alpha)  = \bf{Q}(\alpha) \cdot\nabla\bar{\bf{v}} \cdot \bf{Q}^T(\alpha)$, where $\bf{Q}$ is the rotation matrix. We conduct the following simulation in three stages: \textit{i)} we initially apply a simple shear boundary condition until the systems stabilize, \textit{ii)} sudden rotation ($\alpha=\pi/4$) on the velocity gradient is applied, letting the system evolves over three folds its relaxation time ($\lammic$), and \textit{iii)} the velocity field is suspended to let the systems reach equilibrium no-flow condition. 
\\
\\
In figure \ref{fig:rotation}.\textit{b}-\textit{c}, we present the variation of the mean orientation angle ($\Delta \theta=|\theta - 2(\alpha+\theta^{\parallel})|$) and the mean end-to-end distance ($R_f$) of the oligomer coils. Where $\theta$ is the angle between the end-to-end vector and the $x$ axis (in the fixed reference frame), $\theta^{\parallel}$ is the angle formed by the end-to-end vector and the $\bf{v}$ when $\alpha=0$, and $\theta^o$ is the mean angle under no flow condition. Since at microscopic scales the orientation time can be affected by the thermal fluctuations of the system, in figure \ref{fig:rotation} we compare the coil state for two different temperatures. In general, we identify the rotation in the velocity field effectively induces an affine alignment of the mean orientation angle with the flow, thus as $\alpha$ increases the coils rotate to preserve $\Delta \theta$. Similarly, the measured size of the coils remains unchanged during the sudden rotation of the flow. Therefore, the transition of pure shear to mixed flow does not induce additional stresses on the coils. The reduction of $R_f$ at the final stage of the simulation (under no-flow condition) is evidence that the stretching of the coils is effectively induced by the imposed flow. Additionally, under no-flow condition the mean-angle of the coils converges to $45$ consistent with ramdomly distributed chains (angle averaged over the first quadrant). In figure \ref{fig:rotation}.\textit{b} the coil reorientation response induced by the large sudden change in $\alpha$ occurs on time scales smaller than the microscopic relaxation time $\lammic$. In practice, in an LHMM simulation, large changes in the flow orientation ($\alpha$) are not likely to occur in a single macroscopic time step. Therefore, we expect that orientational relaxation will always occur at time scales smaller than the overall time of a microscopic simulation. 

\subsubsection*{Complex fluid characterization}

\begin{figure}%[!h]%tbhp]
	\centering
	\setlength{\unitlength}{0.1\columnwidth}
	\scalebox{1}{
	{\includegraphics[trim=0cm 0cm 0cm 0cm, clip=true, width=1\textwidth]{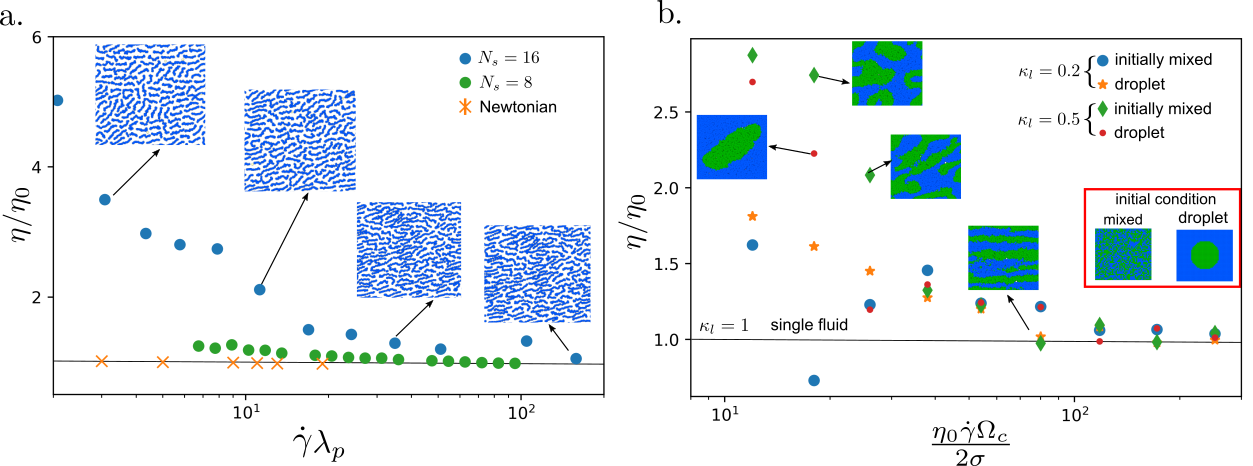}}		
	}	
	\caption{Characteristic viscosity $\eta/\eta_0$ of the complex fluid adopted as a function of the imposed shear rate $\dot{\gamma}$. $\eta_0$ is the input solvent viscosity (or viscosity of the Newtonian fluid), and $\eta$ is the measured total fluid viscosity. a) oligomer melt with $N_s=16$ and b) biphasic flow with two different compositions ($\kappa_l=0.2$ and $\kappa_l=0.5$). Multiphase flow is modelled using two initial condition that lead to different effective viscosities.}
	\label{fig:micrRheo}
\end{figure}

Before proceeding with the validation of the LHMM, we characterize the modelled fluids at the microscale (oligomer melt and multiphase flow) and corroborate that effectively exhibit a complex rheological response. In figure \ref{fig:micrRheo}, we present the response of both oligomer melt and multiphase fluid under simple shear. The oligomer melt exhibits the characteristic shear thinning behaviour, induced by the alignment of the coils in the system as the shear rate increases \citep{Nieto2022}. The relaxation time $\lambda_p$ for the two models of chains used ($N_s=8$ and $N_s=16$) are $\lambda_p \approx 6 t_{\text{SDPD}}$ and  $\lambda_p \approx 9 t_{\text{SDPD}}$.

The flow constituted by two liquid phases ($l$ and $k$) also shows a reduction in the viscosity as the capillary number of the system increases. At the lowest $Ca$ modelled, the low affinity between phases induces the formation of interfaces raising the overall viscosity of the system. As the capillary number increases, the mixing of the phases or alignment is favoured leading the system to the viscosities of the individual phases. For multiphase flow, the characteristic time $\lambda_{\text{ps}}$ of phase separation is a relevant time scale that can determine the stress level of the system. In general, the flow can affect the rate and trajectory of the phase separation leading to metastable microstructures\cite{Minale1997}, or completely inhibiting the phase separation to occur. For comparison, in figure \ref{fig:micrRheo}.b, we consider two different initial conditions \textit{i)} fully phase-separated system, and \textit{ii)} fully mixed phases. In the scenario \textit{(i)} the phase $k$ is modelled as a phase-separated droplet that is subjected to a shear flow. Corresponding to the condition where $\lambda_{\text{ps}}$ has been reached (complete phase separation has occurred). In contrast, in \textit{(ii)} both phases are randomly distributed in the domain when the shear flow is imposed. Thus, the stress evolution of the systems occurs on time scales smaller than $\lambda_{\text{ps}}$. Overall, we observe that at low shear rates the viscosity of the system is strongly related to the extent of the phase separation. Whereas for high shear rates (large $Ca$), the effects of interface formation are significantly reduced, and the system exhibits the characteristic simple-phase viscosity.  In Appendix figure \ref{fig:micrRheoEvol}, we have included the temporal variation of the stress for four different capillary numbers to highlight the differences in the stress evolution for systems undergoing phase separation. In general, multiphase systems with longer relaxation times require a detailed track of their microstructure to ensure an adequate description of the stress and macroscopic flow response. Examples of such complex systems include biological aggregations (cells and proteins), which clusters can extend over several spatial and temporal time scales.  

\subsection{Fully microscopically resolved simulations}

% Before addressing the modelling of more complex physical problems in LHMM, we highlight the significant spatio/temporal gain attainable with the methodology proposed. 

\begin{figure}
	\includegraphics[width=1\textwidth]{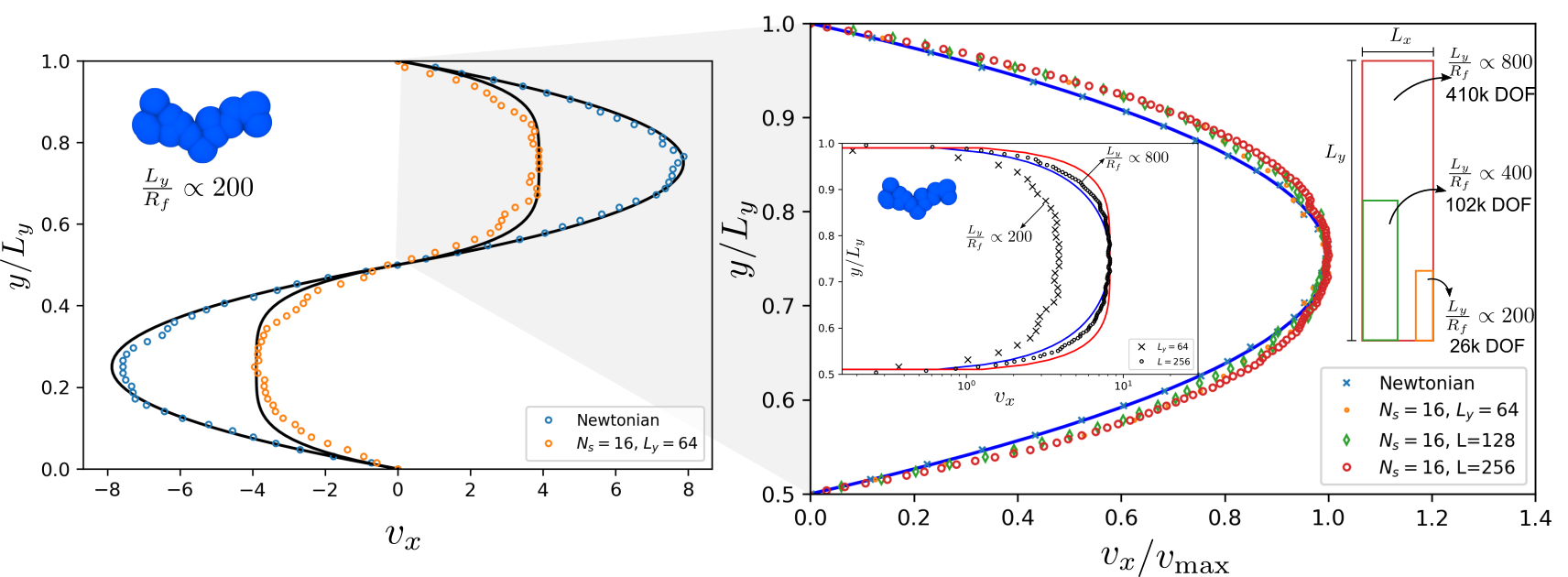}
	\caption{\textit{a}. Fully microscopically resolved RPF for Newtonian and non-Newtonian fluid. Non-Newtonian fluids are modelled oligomers with$N_s=16$ particles per chain. \textit{b}. Closeup of the upper part of a RPF for different domain sizes evidencing a characteristic non-Newtonian profile. We indicate the total number of degrees of freedom (particles) required in those simulations along with the box and oligomer size ratio, for each case. Larger domains will readily required DOF $> 1\times10^6$}
\label{fig:fullMicro}
\end{figure}

To validate the accuracy of the proposed LHMM, we conduct RPF simulations of a fully resolved (micro) Newtonian fluid and oligomeric melts using simulation domains with length-scales on the order of $\bar{L} \propto 10^{2}\dhmi$ to $10^3\dhmi$. For oligomeric melts, we can refer to the domain size in terms of the end-to-end distance of the coils. As a consequence, for oligomers with $N_s = 16$ and $R_f \approx 1.5\dhmi$, the fully resolved domains corresponds to lengths on the order of $200R_f$ to $800R_f$.  These fully resolved systems require between $10^{3}$ to $10^6$ microscopic particles or degrees of freedom (DOF). We must remark that macroscopic domains using fully resolved microscopic scales can be typically on the order of $\bar{L} > 10^{8}\dhmi$, thus requiring $DOF >10^{9}$. The computational cost to simulate such large systems quickly becomes prohibitive, even for efficiently parallelizable codes. The domain size used herein, provides a baseline to evaluate the accuracy of the proposed LHMM framework and is already large enough to evidence the high computational demand for this type of system. In figure \ref{fig:fullMicro}\textit{.a}, we compare the velocity profiles for a fully-resolved Newtonian and an oligomeric melt. Under the same forcing, the non-Newtonian behaviour of the oligomeric melt is evidenced by a reduced velocity (larger viscosity) and flattened profile. Solid lines correspond to the quadratic and fourth-order fitting of the velocities for the Newtonian and melt, respectively. In figure \ref{fig:fullMicro}\textit{.b}, we present the velocity profile of the upper side RPF velocity profile obtained for three different domain sizes with fixed $\nabla \vcon_{xy}|_{\text{max}}$. As the domain size increases the effective velocities of the system change. However, the non-Newtonian profile is consistently preserved.

\begin{figure}
	\includegraphics[width=1\textwidth]{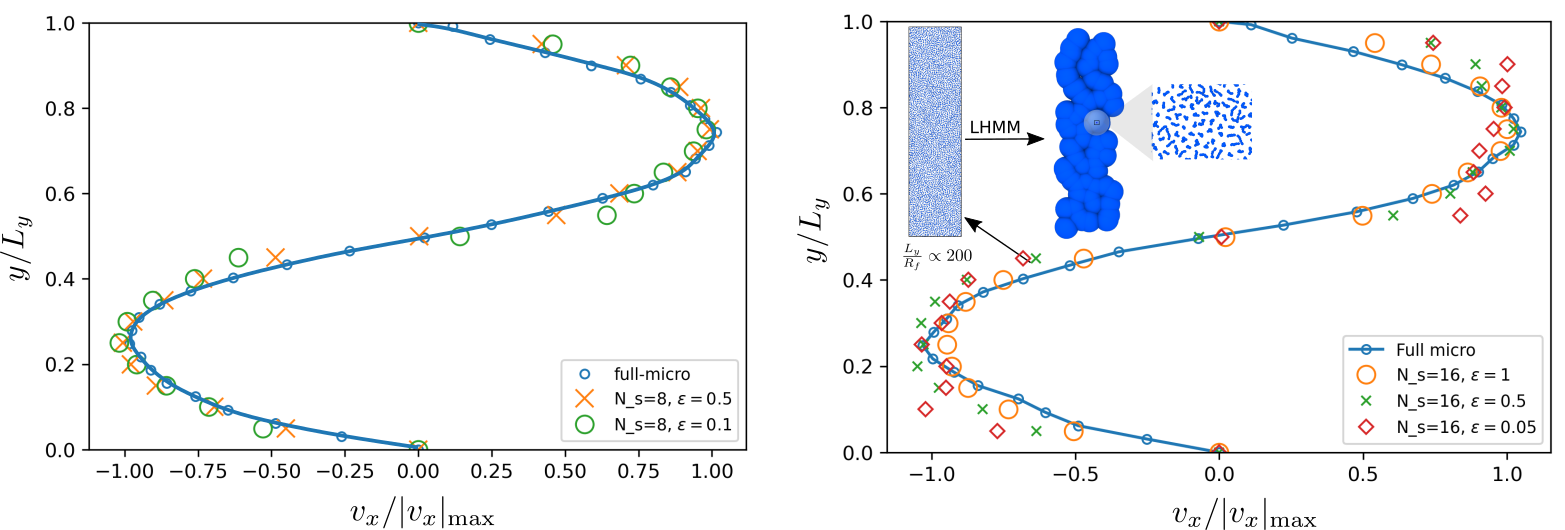}
	\caption{RPF velocity profiles for two different oligomeric melts with a$N_s=8$ and $N_s=16$ using different values of the stabilizing parameter $\epsilon$. Overall, the LHMM schemes captures up to a good approximation the effect of microscopic oligomer chains in the flow. For $N_s=16$, relative larger deviations are observed as $\epsilon$ approximate zero. This is likely originated by the noise-to-signal ratio for in the computed stress for larger chains. Further improvement can be achieved by increasing the sampling volume at the microscales.}
\label{fig:fullMicrovsLHMM}
\end{figure}

In figure \ref{fig:fullMicrovsLHMM}, we compare the corresponding velocity profiles obtained from fully resolved microscopic solutions and the proposed LHMM for two oligomeric systems ($N_s=8$ and $N_s=16$, with $\bar{L}=64$). LHMM results correspond to simulations with $\dhma=3.2$ and $\dhmi=0.2$. Considering a kernel size $\bar{h} =4\dhma$ and a microscopic domain $\omes = 20\dhmi$, the spatial gain for these test is $G_s=3.2$. We evaluate the influence of the stabilizing parameter $\epsilon$. In general, we observe that when hydrodynamic contributions are only accounted for from macro simulations $\epsilon \approx 1$ the effective viscosity of the systems increases leading to slightly smaller velocities for LHMM. Such effect is reduced as $\epsilon$ diminishes. When hydrodynamic $\epsilon \approx 0$ the obtained velocity exhibit instabilities, that are likely related to the macroscopic particle resolution. Since the stresses are only accounted for from microsimulations when $\epsilon = 0$, the viscous interactions between macroscopic particles can experience numerical fluctuations due to the stress calculation from microscopic transient simulations. However, we must note that even for $\epsilon = 0$ the order of magnitude of the viscous stresses is closely related to the fully microscopic results.
LHMM with $\epsilon \approx 0$ reproduces up to a good approximation the characteristic behaviour of the oligomeric system. Overall, we identify that stabilization parameters $\epsilon>0.1$ provide a reasonable stabilization of the stresses.   

\subsection{LHMM for complex fluids}

Now, we continue evaluating the proposed LHMM on a macroscopic domain with significantly larger spatio/temporal gain, solving the microscales using the SDPD equations \eqref{eq:possdpd} to \eqref{eq:random}. For macroscopic simulations we consider a fluid with properties $\rho=1000$ $\si{Kg/m^3}$, $\bar{\eta}=1e-3$ $\si{Pa \cdot s}$, $c=0.1$ $\si{m/s}$, $\bar{p}_b=1$ $\si{Pa}$. The macroscopic time and length scales are defined in terms of $\Delta \bar{t}=0.0002\si{s}$, $\dhma = 5\cdot10^{-4}\si{m}$, and $\bar{h}=0.002\si{m}$ respectively (see table \ref{tab:macro}).  For microscales, we adopt a resolution $\dhmi = 2.5\cdot10^{-10}\si{m}$, such that the size of the microscopic kernel is ${h}'=4\dhmi=1\cdot10^{-10}\si{m}$, and $\omes = 20\dhmi$. Therefore, these LHMM simulations correspond to spatial gains $G_s \approx 4\cdot10^{6}$. From \ref{eq:Nlhmm}, we can observe, that it implies a reduction in the required DOF of $\propto 10^{6}$, compared to the fully-microscopically resolved system. 

\begin{table}
  \begin{center}
\def~{\hphantom{0}}
  \begin{tabular}{lcc}
      parameter  & symbol   &   value \\[3pt]
      box size  & $L$ & $1\si{mm}$ \\
      density & $\rho$ & $1000 \si{Kg/m^3}$ \\
      viscosity & $\bar{\eta}$ & $1e-3$ $\si{Pa \cdot s}$\\
      bulk viscosity & $\zeta'$ & $3/5\bar{\eta}$ \\
      speed of sound & $c$ & $0.1 \si{m/s}$\\
      background pressure & $\bar{p}_b$ & $1\si{Pa}$ \\
      resolution & $\dhma$ & $5\cdot10^{-4}\si{m}$ \\
      kernel size  & $h$ & $4\dhmi$ \\
      time step & $\Delta \bar{t}$ & $0.0002\si{s}$
  \end{tabular}
  \caption{Macroscopic fluid and system parameters}
  \label{tab:macro}
  \end{center}
%   \label{tab:macro}
\end{table}

To streamline the construction of the different microscopic systems and presentation of the results, we conduct microscopic simulations using reduced units ($X'=X_{\text{physical}}'/X_{\text{ref}}'$). We introduce a reference length ($h_{\text{ref}}'= 1.25\cdot10^{-9}\si{m}$), mass ($m_{\text{ref}}'= 1.56\cdot10^{-15}\si{Kg}$), and time ($t_{\text{ref}}'= 0.0125\si{s}$) scales (see table \ref{tab:micro}). Henceforth, unless otherwise stated, the reduced fluid properties of the microscopic simulations are consistently given by $\rho'=1.0$, ${\eta}'=10$. The particles are initially localized in a square grid with an interparticle distance $\dhmi= 0.2$. Additionally, for microscopic simulations we use $c'=40$ and ${p}_b'=50$. The time step is chosen to satisfy the incompressibility of the system and ensure numerical stability, we choose the smaller time scale between the Courant-Friedrichs-Lewy condition $\Delta t' = h'/(4(\vcon +c)$ and the viscous time scales, $\Delta t' = h'^2/(8\eta'\rho')$. Thus, we use $\Delta {t}'=0.0001$ to ensure lower density fluctuations. At microscales, we account for thermal fluctuation, thus the energy scale is determined by  $k_BT=1.0$. Following the results reported by \citep{Moreno2021}, we construct microscale simulations suitable for arbitrary boundary conditions with a core size between $\omes  = 15\dhmi$ and $\omes  = 20\dhmi$ (i.e. the size of the sample to determine the stresses), whereas the sizes of the boundary condition and buffer regions are $5\dhmi$ and $5\dhmi$, respectively. 

\begin{table}
  \begin{center}
\def~{\hphantom{0}}
  \begin{tabular}{lcc}
      parameter  & symbol   &   value \\[3pt]
      physical resolution & $\dhmi$ & $2.5\cdot10^{-10}\si{m}$ \\
      physical kernel size  & $h$ & $4\dhmi$ \\
      reference lenght scale & $h_{\text{ref}}'$ & $1.25\cdot10^{-9}\si{m}$ \\
      reference mass scale & $m_{\text{ref}}'$ & $1.56\cdot10^{-15}\si{Kg}$ \\
      reference time scale & $t_{\text{ref}}'$ & $0.0125\si{s}$ \\
      density & $\rho'$ & $1$ \\
      viscosity & $\eta'$ & $10$\\
      bulk viscosity & $\zeta'$ & $3.5\eta'$ \\
      speed of sound & $c$ & $40$\\
      background pressure & $p_b'$ & $50$ \\
      resolution & $\dhmi$ & $0.2$ \\
      kernel size  & $h$ & $4\dhmi$ \\
      time step & $\Delta t'$ & $0.0001$ \\
      core size  & $\omes$ & $15\dhmi$  to $20\dhmi$ \\
      buffer size & $\obuf$ & $5\dhmi$ \\
      bc size & $\obc$ & $5\dhmi$
  \end{tabular}
  \caption{Microscopic fluid and system parameters}
  \label{tab:micro}
  \end{center}
%   \label{tab:micro}
\end{table}

In Appendix figure \ref{fig:vgradvEpsMicro}, we compile the steady state velocity profiles obtained for a Newtonian flow, using the stress tensor computed  directly from microscopic subsystems (\eqref{eq:skin} and \eqref{eq:spot}), for various $\epsilon$. Consistently, in figure \ref{fig:vgradvEpsMicro} we can observe that micro-scale simulations can recover the macroscopic stress tensor, leading to the proper modelling of the flow. This represents an evidence of the robustness of SDPD to capture the ideal solvent contributions across scales. For comparison, we have included the velocity profile obtained for a system without microscale contributions.

% However, compared with the simulated LHMM systems, these fully resolved domains are one order of magnitude smaller. 
% Thus evidencing the significant spatio/temporal gain attained. 
% 

\subsubsection{Oligomeric melt}

\begin{figure}
	\includegraphics[width=0.8\textwidth]{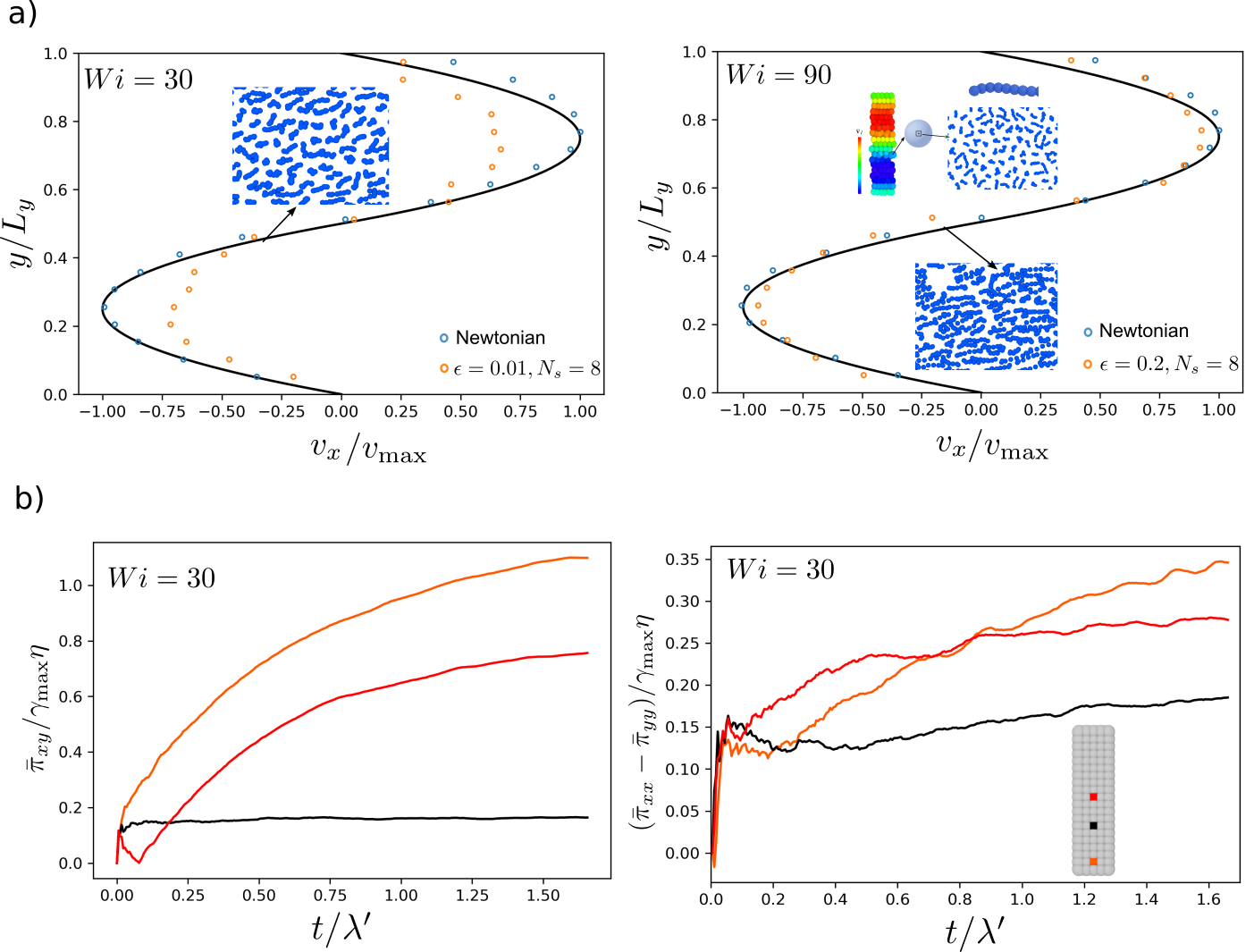}
	\caption{RPF for non-Newtonian fluid, using oligomers with $N_s=8$. $a.$ Comparison between Newtonian and non-Newtonian for two values of $\epsilon = 0.01$ and $\epsilon = 1.0$. The magnitude of the stabilization parameter in do not affect the rheology of the fluid, evidencing a proper description of the fluid from both macro and micro scales separately.  Shear-thinning effect fo oligomer melts at different shear rates. The value of $\epsilon=0.01$ is used for low shear rates, as the shear rate increases the value of $\epsilon=0.2$ is used to ensures stability of the measured stress. $b)$ Variation of the stress and first normal stress differences for three different macroscopic particles. The initial position of the particles is highlighted on the right. Microscales constituted by oligomers with $N_s=8$. Three different RPF configurations varying the velocity gradients are compared. The characteristic normal stress differences in the fluid increses due to the microscopic response of the chains.}
\label{fig:polyLHMM}
\end{figure}

In figure \ref{fig:polyLHMM}, we show the steady state results for a RPF flow configuration of oligomeric melts at different shear rates. We can observe the characteristic shear-thinning effect induced by the alignment of the chains in the flow. Overall, the magnitude of the stabilization parameter did not induce any effect on the rheology of the fluid, evidencing a proper description of the fluid from both macro and micro scales separately. In addition to the steady state solution, we were able to capture the characteristic deviations in the temporal evolution for the oligomer melt (see Appendix). For the Newtonian fluid, the velocity profile is consistently reproduced by a quadratic fitting, whereas the microscopic effects of the oligomer chains lead to a 4th order velocity profile in the non-Newtonian fluid.

% In figure \ref{fig:polyLHMM}.b we have included the mean orientation angle and end-to-end distance of the oligomers for the different shear rates to evidence such alignment effect in the microscales.

% In addition to the shear-thinning of the oligomeric melt, another relevant feature that can 
Besides the differences in the velocity profile for oligomeric melts, another relevant characteristic that can be analysed for this non-Newtonian fluid is the evolution of their stresses. In figure \ref{fig:polyLHMM}.b, we present the variation of the shear stress and first normal stress difference for three macroscopic particles (highlighted in red, black and orange) at $Wi=30$, for oligomer melts with $N_s=8$. The particles are initially localized at positions across the domain such that they experienced different magnitudes of stress. As described in figure \ref{fig:polyLHMM}, a shear-thinning behaviour can be evidenced in the magnitude of $\bar{\pi}_{xy}$ when the shear rate increases. Additionally, the emergence of first normal stress differences is observed for the macroscopic particles due to the microscopic response of the chains. 
% Even for this simple RPF, the feature that every particle carries its stress history is evident for $\nabla \bar{v}_{xy}|_{\text{max}}=16$, where the \textit{orange} particle (initially located in the bottom) changes its vertical position to a region with lower shear rate. This lead to a reduction in the particle stress and normal stress differences. However, the change does not occur immediately but entails a continuous transition of the particle from its original state.     

% \begin{figure}
% 	\includegraphics[width=1\textwidth]{./tauandN1Poly}
% 	\caption{Variation of the stress and first normal stress differences for three different macroscopic particles. The initial position of the particles is highlighted on the right.  
% 	microscales constituted by oligomers with $N_s=8$. Three different RPF configurations varying the velocity gradients are compared. The characteristic normal stress differences in the fluid increses due to the microscopic response of the chains.}
% \label{fig:stressAndN1}
% \end{figure}

\subsubsection{Multiphase flow}
\begin{figure}
\centering
	\includegraphics[width=0.9\textwidth]{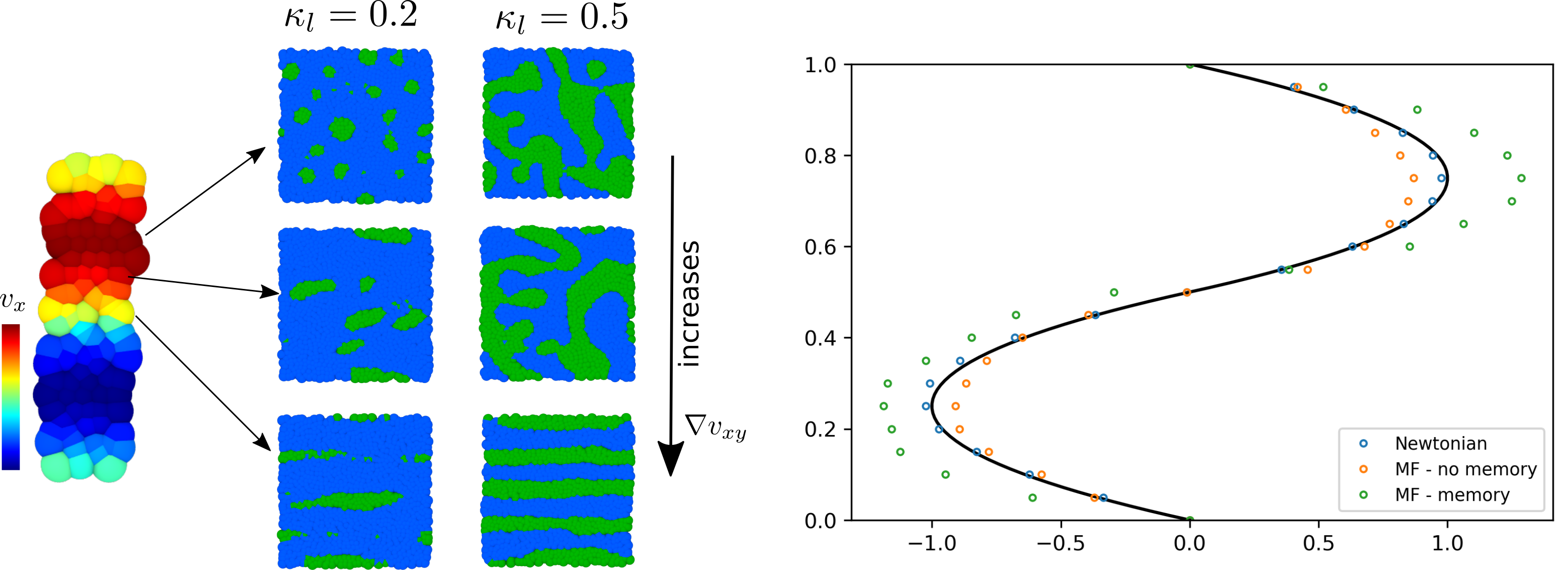}
	\caption{Typical velocity profile for RPF coupled with multiphase flow at microscale, using two different compositions of the phase $l$, $\kappa_l=0.2$ and $\kappa_l=0.5$. In \text{b} we compare the steady-state velocity profile obtained for a Newtonian fluid and two schemes of LHMM simulations. The phase separation at microscales originates a shear-thinning behaviour of the macroscopic flow. For comparison, we include the steady profile for a system without historical tracking of the microscale. In that situation, the formation of microstructures with larger relaxation times is never reached, and the fluid behaves similar to the Newtonian fluid.}
	\label{fig:mfFlow}
\end{figure}

Using the same RPF setting at the macroscales, we can easily investigate other physical systems with different microscopic features. In figure \ref{fig:mfFlow}, we compile the results obtained for multiphase flows using two phases $l$ and $k$, with compositions $\kappa_l=0.1$ and $\kappa_l=0.5$. In these simulations the two phases are initiallity mixed and the phase separation takes places concurrently with the imposition of the flow. As a result, the macroscopic shear affects the morphology of the microstructure formed, leading to a different response of the mixture. The characteristic size of the microstructure depends on the phase composition. Low concentrations of $l$ phase favour spherical to elongated droplet transitions, whereas at intermediate concentrations the increase in the shear rate induces transitions of the microstructure from disordered spinodal to lamella-like structures. 
\\
\\
In figure \ref{fig:mfFlow}, we also compare the steady-state velocity profile obtained for a Newtonian fluid and the multiphase case with $\kappa_l=0.5$. For comparison, we have included the profile for a multiphase system where the microstructure at the begining of each macroscopic time step is reinitialize as fully mixed. This assumption is consistent with microstructural evolution reaching its equilibrium condition on time scales much smaller than the macroscopic time step. However, for these type of system it will imply that the historical evolution of the microstructure is neglected. In general, we observe that the microphase separation originates a shear-thinning behaviour for the multiphase systems modelled. Remarkably, we can observe that the thinning behaviour roots in the proper history tracking of the microstructure. In systems without memory, the formation of microstructures with larger relaxation times is never reached, and the fluid resembles the Newtonian behaviour of their individual phases.

\subsection{Flow through complex geometry}

Now, we evaluate the proposed LHMM framework on geometries that induces different local flow types (i.e shear, extension, and mixed flow), for both oligomer melts and multiphase flows. For these large macroscopic domains, a direct validation with the fully microscopically resolved systems is computationally taxing. Therefore, for complex geometries, we first validated the simple Newtonian fluid in the LHMM scheme, with respect to the corresponding Newtonian fluid as modelled from a macroscopic simulations (using only SPH simulations) (see Appendix figure \ref{fig:newFlowCylLHMM_Ma}). Overall, we identify that the LHMM consistently captures the behavior of the ideal fluid, on the range of paramters evaluated.
% The use of arbitrary BC at the microscale allows us to account for different spatial flow configurations. In figure \ref{fig:complexG} we present the two types of geometries investigated, corresponding to a flow passing a circular and square contraction. We model the walls ensuring 

\subsubsection{Oligomeric melt}

\begin{figure}
\centering
\includegraphics[width=0.9\textwidth]{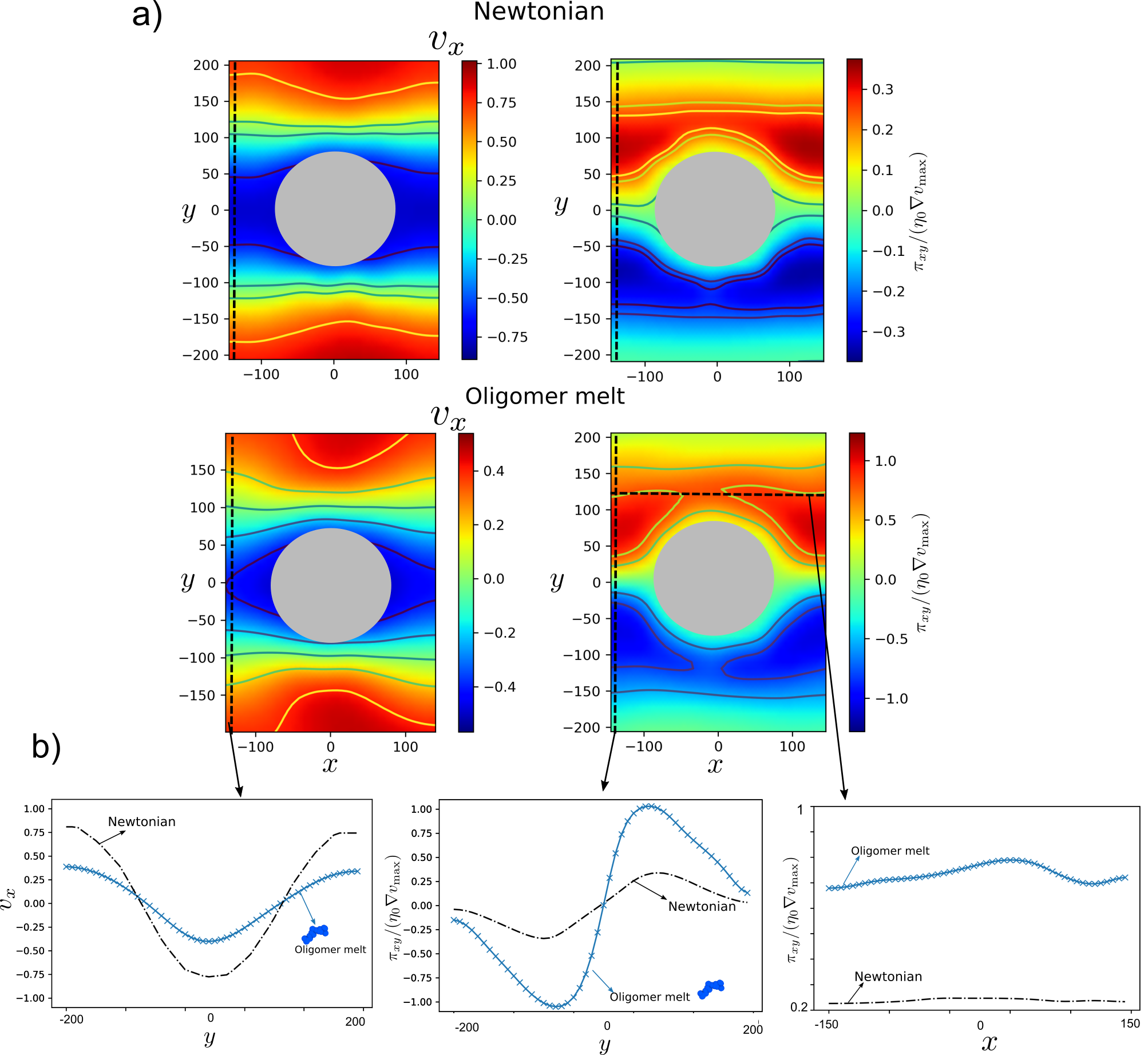}
\caption{Comparison of the velocity and hydrodynamic stresses contours (a)) between a Newtonian and oligomeric melt around a cylinder (at $Wi=0.4$). In a domain of size $19\dhma \times 27\dhma$ with a radius of cylinder of $R=6\dhma$.  b) Comparison of the velocity profile along a vertical line at the entrance of the channel. Microscopic features of the chains at the microscales induce the deviation of the Newtonian behaviour leading flattened velocity profile. Stress along a vertical and horizonal lines are presented to compare both fluids.}
\label{fig:cavities}            
\end{figure}

\com{In figure \ref{fig:cavities}, we present the steady velocity and stresses for an oligomeric melt ($N_s=16$), passing a cylindrical array at $Wi=0.4$. For the cylindrical contraction we use a domain of size $19\dhma \times 27\dhma$, and the radius of the cylinder $R=6\dhma$. The size of the macroscopic kernel $4\dhma= 64$ and the microscopic domain size $\Omega'=8$ are defined such that the overall spatial gain of these simulations is $G_s=8$, and an aspect ratio between the cylinder $R$ and the coil size $R_f$ of nearly $300$ times. The flow at the macroscale is induced by an external forcing $f_{\text{ext}}=0.58$, acting on the fluid particles. Fully microscopically resolved simulations of these systems would require over $10^7$ particles for a two-dimensional system, in contrast to the $10^6$ particles used for LHMM. Figure \ref{fig:cavities}.a compares the velocity and stress contours between a Newtonian and oligomeric melt. In general, we identify alterations in the steady profiles arising from the enhanced viscosity of the oligomer melts. The characteristic shear thinning response of the melt (as discussed in previous sections) to the spatially-changing velocity gradient induces a modest but evident break in symmetry for both velocity and stress. In figure \ref{fig:cavities}.b, we plot the profiles along the vertical line at the entrance of the domain. The higher viscosity of the oligomeric melts is consistent with the typical flattened velocity profile observed and the larger stress. The stress profiles along a vertical and horizontal lines is also presented in figure \ref{fig:cavities}.b to illustrate the larger stress contribution due to the oligomeric chains and the change in the generated stress along the channel.}

% \subsubsection{Flow passing a square contraction}

% \begin{figure}
% \centering
% 	\includegraphics[width=0.8\textwidth]{./sqCavity}
% 	\caption{Velocity profile for a polymer solution around a square contraction with two different polymer concentration. Polymer concetration is considered homogeneous in the whole domain {{\color{red} PENDING RESULTS ON HMM, AND BETTER CONTOURS}}}
% 	\label{fig:sqCavity}
% \end{figure}

\subsubsection{Multiphase flow}

\begin{figure}
\centering
	\includegraphics[width=0.9\textwidth]{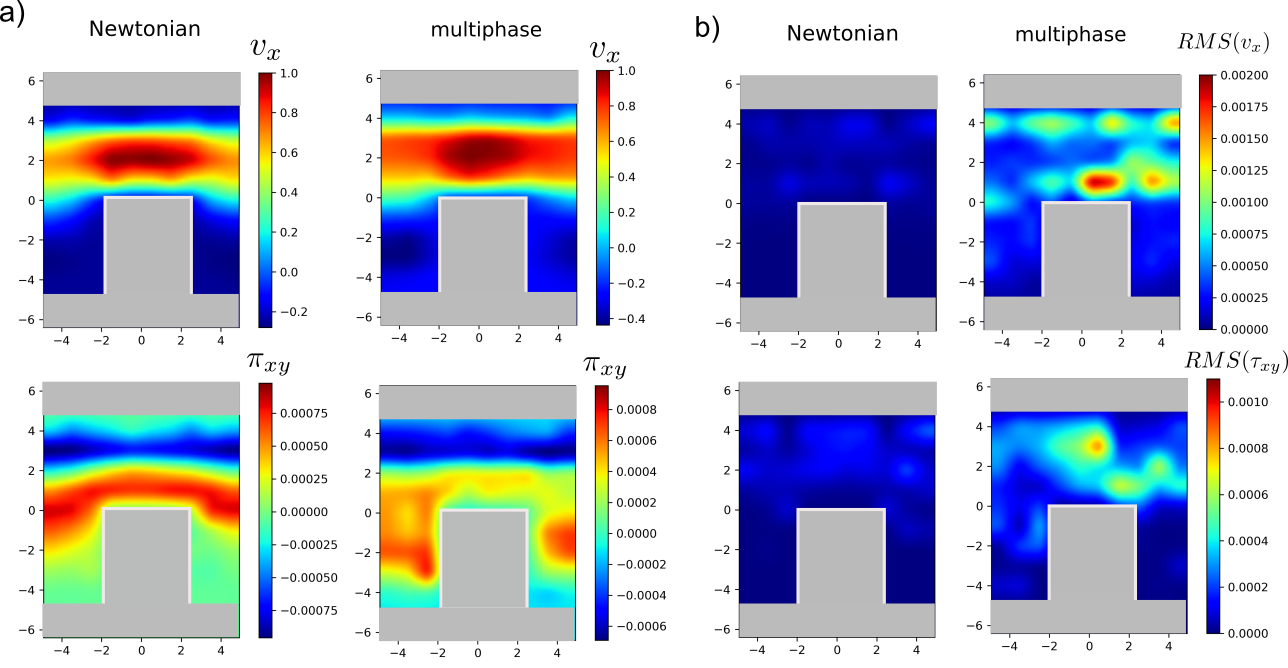}
	\caption{Comparison of the velocity and stress $\bar{\pi} _{xy}$ on a square-contraction array for Newtonian and multiphase flows. \textit{a} steady-state velocity and stress contours \textit{b} mean root squared of the velocity and stress fluctuations}
	\label{fig:trajStress}
\end{figure}

\com{The capabilities of the method to track history-dependent effects are further shown using multiphase flows in a square cavity array. In figure \ref{fig:trajStress}, we include the velocity and stress profile for a Newtonian and a biphasic fluid,  averaged over the same macroscopic time span.
For the biphasic fluid the microscale simulations are initialized as homogeneously mixed phases, with $\kappa_k=0.5$, that undergo microphase separation as they flow through the channel. As a result,  multiphase flows are characterized by the emergence of microstructures that can evolve with the simulation, therefore carring historical information during their transport across the channel. The formation of such microstructures is additionally affected by the spatially-variable velocity gradient experienced by each macroscopic particle. Consistently, as the particles move within the domain the state of the microstructure determines their stress response, affecting the macroscopic flow. In \ref{fig:trajStress}.a, we can observe that the Newtonian fluid has reached a nearly symmetric steady condition for both velocity and stresses. Whereas the multiphase fluid exhibit a significantly different flow behavior and stress distribution. Further estimation of the root-mean-squared (RMS) of the velocity and stresses fluctuation allows us to elucite that the temporal stability of the velocity and stress are responsible for the observed flow patterns. Figure \ref{fig:trajStress}.b evidences the persistent fluctuations on multiphasic systems, due to the continuous evolution of the microstructure. Different from simple Newtonian fluids, multiphase flows are likely to require larger simulation times in order to reach an steady-state condition (in the statistical sense). In Appendix figure \ref{fig:mfevol}, we present the evolution of the velocity and stress (and RMS of the fluctuations) for multiphase flow at different time steps, evidencing that multiphasic systems have not reached yet an steady condition. We must, highlight that when comparing the stress evolution between Newtonian and multiphase flows, the latter is characterized by larger relaxation times ($\lambda_{ps}$) that typically exceed a single microscopic simulation. The LHMM used herein, allows us to naturally account for such large relaxation times while keeping the modelling of microscopic simulations computationally feasible. } 

It is important to note that depending on the characteristic size $r_{\text{micro}}$ of the microstructure, the size of the microscopic domain must be large enough for the microstructure to be commensurated, this is $\Omega'> r_{\text{micro}}$. Since certain physical systems can exhibit microstructures constantly varying in size (e.g. continuously growing aggregates), the definition of $\Omega'$ poses some important challenges, requiring a systematic analysis of the specific physical phenomena investigated. However, these aspects related to varying microstructural size are out of the scope of the present work and will be addressed in future publications. Here, we have focused on showcasing the capabilities and flexibility of the proposed approach. 

% by the presence of different local flow types: simple shear in regions that are sufficiently far from the ends

% of the contraction, extensional and mixed motion in proximity of entrance and exit of the contraction, and

% mixed rotational flow near the corners of the domain. .

\section{Conclusions and future work}
Herein, we proposed a fully-Lagrangian Heterogeneous Multiscale Methodology, suitable to model complex-fluids across large spatial/temporal scales using fluctuating Navier-Stokes equations.  This methodology offers the advantage of capturing microscopic effects at the macroscopic length scales, with a lower cost than solving the full microscale problem in the whole domain. The LHMM discretize both macro and microscale using the smoothed dissipative particle dynamics method, taking advantage of its thermodynamic consistency and GENERIC compliance. The LHMM uses the velocity field of the macro scales to define the boundary conditions of microscale subsystems that are localized at the positions of the macroscopic particles. Subsequently, those microscale subsystems provide a microscopically derived stress $\bm \tau$ that is pushed to the macro scales to close the momentum equation and continue their temporal solution. This way, the stress information is explicitly carried by the macroscopic Lagrangian points and memory effects related to the evolution of the microstructure are preserved. The microscale domains can be constructed on-the-fly, wherever they are required, based on the evolution of the macro simulation or at prescribed intervals to obtain microscale-informed properties. We tested the LHMM using both Newtonian and non-Newtonian fluids evidencing its capability to capture complex fluid behaviour such as polymer melts and multiphase flows under complex geometries. The LHMM was developed using the highly parallelizable LAMMPS libraries. An important feature is that both macro and microscale can be fully parallelized separately. This has significant advantages compared to fully microscopically resolved systems, which required intensive communication between subdomains of the system. In LHMM, each microscopic simulation is executed separately reducing communication bottlenecks. Further, applications of the LHMM  include various complex systems such as colloidal suspensions or biological flows.

% {\color{red} PENDING FURTHER DISCUSSION}

\pagebreak

% \backsection[Supplementary data]{\label{SupMat}Supplementary material and movies are available at \\https://doi.org/10.1017/jfm.2019...}

\textbf{Acknowledgements:} {The authors acknowledge the financial support received from the Basque Business
Development Agency under ELKARTEK 2022 programme (KAIROS project: grant KK-2022/00052).
Financial support received from the Basque Government through the BERC 2018-2021 program, by the Spanish State Research Agency
through BCAM Severo Ochoa excellence accreditation (SEV-2017-0718) and
through the project PID2020-117080RB-C55 (“Microscopic foundations of soft-
matter experiments: computational nano-hydrodynamics”) funded by AEI - MICIN and acronym “Compu-Nano-Hydro” are also gratefully acknowledged.
N.M acknowledges the support from the European Union’s Horizon 2020 under the Marie Skłodowska-Curie Individual Fellowships grant 101021893, with
acronym ViBRheo.
}

%\backsection[Funding]{This work was supported by}

\appendix
\section{Algorithm of LHMM}

Algorithm of the proposed Lagrangian heterogeneous multiscale method. The core of the algorithm consist of the coupling loop and the calculation of macro and micro scales separately. Both macro and microscales can be parallelizable independently.

\begin{algorithm}[H]
\SetAlgoLined\LinesNotNumbered \DontPrintSemicolon
\SetKwFunction{macro}{macroscales}\SetKwFunction{micro}{microscales}\SetKwFunction{coupling}{coupling}
\SetKwProg{lhmm}{LHMM}{}{}
\lhmm{\coupling{}}{
  \tcc{define set of parameters for MS simulations}
 $\mathcal{N}$ \tcp{number of microscale simulation}
 $\lambda'$ \tcp{total time steps at microscales}
 $\bar{\lambda}$ \tcp{time frequency for sampling the stress}
 $\omes$ \tcp{domain size for microscales}
 $\ocon$ \tcp{domain size for macroscales}
 $\epsilon$ \tcp{stabilizing parameter for macroscopic viscous contributions}
 
 $\bar{\bm \pi} = {\bm 0}$ \tcp{initiallize stress tensor}
 \For{$\bar{t}=0$ \KwTo $\bar{t} = t_{\text{total}}$}{
 \macro{$\Delta \bar{t}$, $\bar{\lambda}$, $\bar{\bm \pi}$,$\epsilon$} \;
\For{$I=0$ \KwTo $I=\mathcal{N}$} {
 Retrieve  $\nabla \vcon_I$\; 
 $\bar{\bm \pi}_I$ = \micro{$\Delta {t}'$, $\lambda'$, $\nabla \vcon_I$}\;
}
 %$\bar{t} = \bar{t} +\bar{\lambda}$ \;
 }
}
\setcounter{AlgoLine}{0}
\tcc{Macroscale simulation}
\SetKwFunction{sdpd}{SDPD}\SetKwFunction{ik}{Irving-Kirkwood}
\SetKwProg{mac}{Macroscales}{}{} \SetKwProg{mic}{Microscales}{}{}
\mac{\macro{$\Delta \bar{t}$, $\bar{\lambda}$, $\bar{\bm \pi}$,$\epsilon$}}{
 \For{$\bar{t}$ \KwTo $\bar{t}=\bar{\lambda}$}{
   \For{I,J $\in$ $\bar{N}$}{
       \sdpd{$\bar{\bm \pi}$,$\epsilon$}\; 
         $\nabla \vcon_I = \sum_J \fijcon \rbijcon \vcon_I$ \;
    }
   $\bar{t} = \bar{t} + \Delta \bar{t}$\;
  }
}
\setcounter{AlgoLine}{0}
\tcc{Microscales simulations with arbitrary-boundary conditions approach}
\mic{\micro{$\nabla \vcon_I$, $\omes$}}{
\For{$t'=0$ \KwTo $t'=\lambda'$}{
   \For{$i,j$ $\in$ ${N}'$}{
   \tcp{the velocities of particles at $\obc$ is prescribed from macro}
      \For{$i$ $\in$ $\obc$}{$\vmes = {\bm r}_i' \nabla \vcon_I$} 
      \sdpd{$\obc$}\; 
    }
     $\bar{\bm \pi}_I$ = \ik{} \;
   $t' = t' +\Delta t'$ \;
  }
  \KwRet{$\bar{\bm \pi}_I$}
}

 \caption{Lagrangian heterogeneous multiscale coupling}
\label{alg:lhmm}
\end{algorithm}

\section{Approximation to the ideal stress}

The ideal stress ($\pima_I^o$) contribution of the hydrodynamic interactions can be computed in different ways. Since the ideal stress tensor from microscales corresponds to the fluid in absence of non-ideal and non-hydrodynamic effects, an alternative is to conduct microscale simulations for a simple fluid at the velocity-field conditions of the particle $I$, and directly compute $\pima_I^o$. However, this approach entails a two-fold increase in the computational cost, requiring keeping track of two microscale systems per each macro particle. Another alternative is to obtain an estimate of $\bar{\textbf{d}}$ using the projection of the macroscopic velocity gradient ($\nabla \vcon_I$) at the microscale 
%
% 
% \begin{align}
%  \nabla \bar{\vcon}_i = \sum_j \left(\vmes_j - \vmes_i)\otimes \textbf{r}_{ij}F_{ij}
% \end{align}
%
%
\begin{align}
\left\langle \nabla \bar{\vcon}_i \right\rangle= \sum_j \left(({\bm r}_j' \nabla \vcon_I - {\bm r}_i' \nabla \vcon_I)\otimes \textbf{r}_{ij}'F_{ij} \right).
\label{eq:projgdv}
 \end{align}
Thus, using the projection \eqref{eq:projgdv} the rate-of-strain tensor $\bar{\textbf{d}}$ can be estimated leading to an ideal stress of the form 
%
% \begin{align}
% \pima_I^P ({\bf x},t) = 2\eta \frac{1}{N_t'}\sum\limits_{n=1}^{N_t'} \frac{1}{2}\left[\sum\limits_{\substack{i,j \\ i\neq j}} \Big(\langle v'_{j}\rangle-\langle v'_{i}\rangle\Big) \otimes \textbf{e}_{ij} m'_{j} w_{IK}({\bf r}_i(n)-{\bf x})\\
% & + \sum_{j}\Big[\Big(\langle v'_{j}\rangle-\langle v'_{i}\rangle\Big) \otimes \textbf{e}_{ij}\Big]^{T} m'_{j} w_{IK}({\bf r}_i(n)-{\bf x})\right]
% \end{align}
\begin{align}
\pima_I^o ({\bf x},t) = 2\eta \frac{1}{N_t'}\sum\limits_{n=1}^{N_t'} \left[\sum\limits_{i} \frac{1}{2} \Big(\left\langle  \nabla \bar{\vcon}_{i}(n) \right\rangle +\left\langle \nabla \bar{\vcon}_{i}(n)\right\rangle^{T} \Big) w_{IK}({\bf r}_i(n)-{\bf x}) \right]
\end{align}

\section{Temporal evolution of stress in binary mixture}

The evolution of the stress for binary systems with different initial conditions. Lower capillary numbers, where interfacial interactions play an important role lead to different stresses as the system evolve. Thus memory effects of the fuid are relevant to properly account for the correct flow behavior. In contrast, systems with larger $Ca$ exhibit similar stress trajectories independently of their initial state.

 \begin{figure}%tbhp]
	\centering
	{\includegraphics[trim=0cm 0cm 0cm 0cm, clip=true, width=0.8\textwidth]{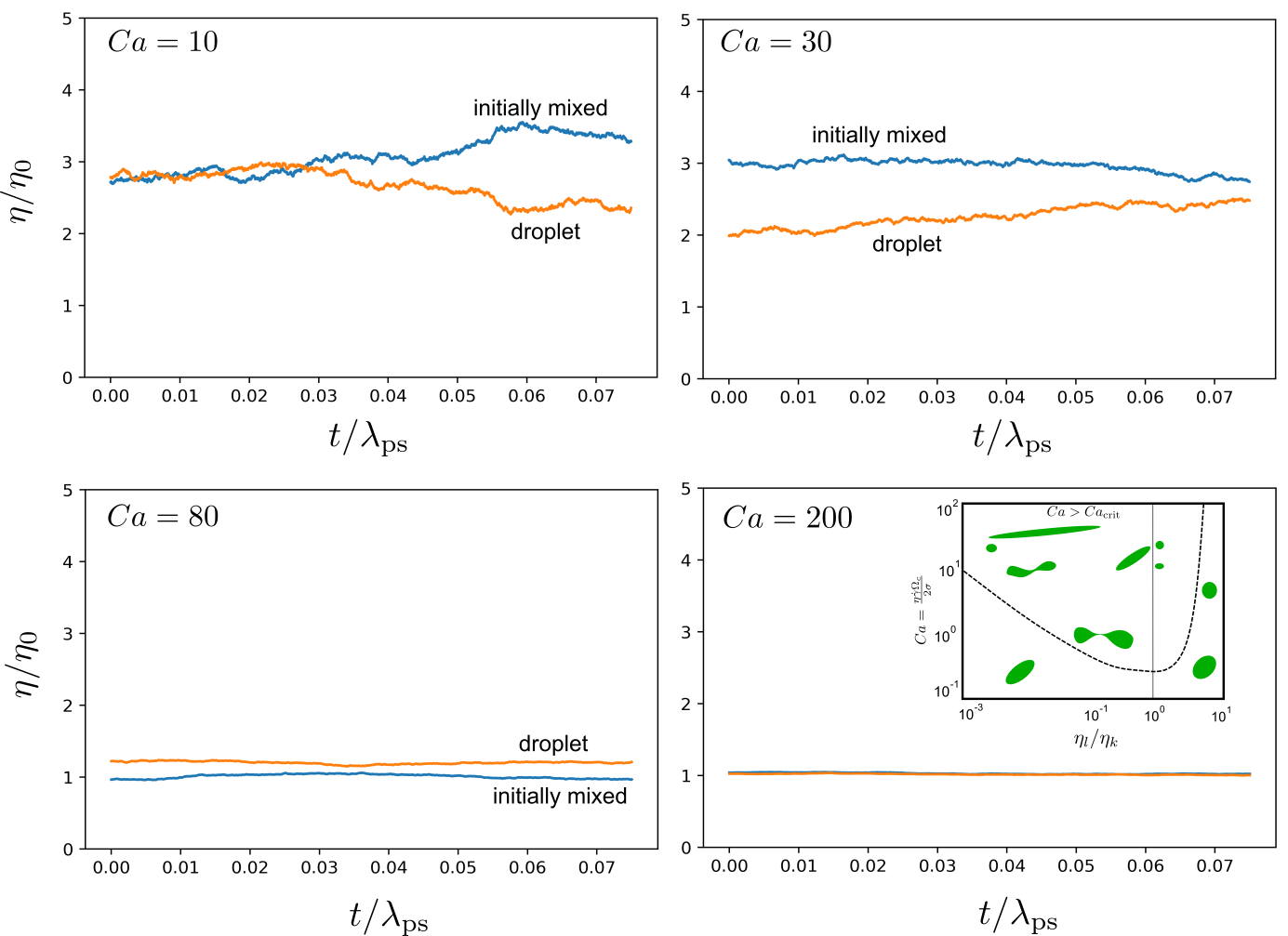}}		
	\caption{Temporal evolution of the microscopic stress for a binary system with $\kappa_l=0.5$, for four different shear rates. Systems with initial condition as fully mixed phases (blue) completely phase separated (orange) are compared. $\lambda_{\text{ps}}$ is the characteristic time for full phase separation to occur. At lower shear rates the effect effect of microstructure formation can affect the effective stress measured. In contrast, for large shear rates both systems exhibit similar shear thinning behaviour, independent of their initial condition. }
	\label{fig:micrRheoEvol}
\end{figure}

\section{LHHM validation for Newtonian fluid}

\begin{figure}%[h!]%tbhp]
	\centering
	{\includegraphics[trim=0cm 0cm 0cm 0cm, clip=true, width=0.7\textwidth]{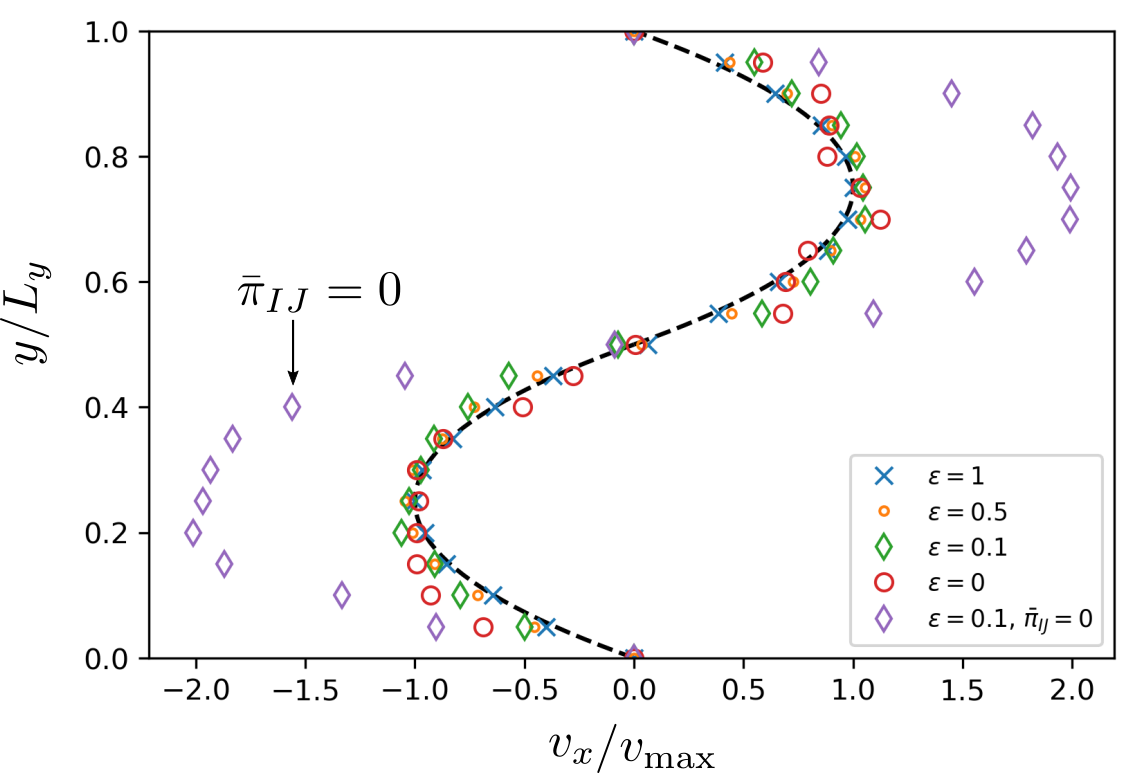}}		
	\caption{Imposed velocity field for macroscales for different values of $\epsilon$ and using the proposed micro-macro coupling. As a comparison, a system without microscales stress cannot recover the desired velocity profile.}
	\label{fig:vgradvEpsMicro}
\end{figure}

\begin{figure}%[!tbhp]%tbhp]
	\centering
	{\includegraphics[trim=0cm 0cm 0cm 0cm, clip=true, width=0.7\textwidth]{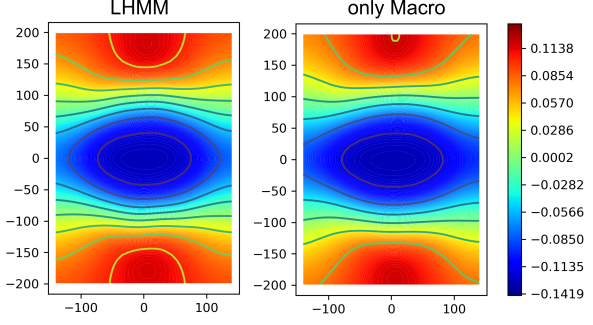}}		
	\caption{Flow arround cylinder LHMM and full macro.}
	\label{fig:newFlowCylLHMM_Ma}
\end{figure}

Effect of the stabilization parameter $\epsilon$ for Newtonian fluid. The LHMM is able to recover the behavior of the fluid upto a good approximation over the whole range of $\epsilon$ investigated. We highlight that the contribution of microscales is fundamental to model the fluid properly. For comparison, in figure \ref{fig:vgradvEpsMicro}, we present the results  for a RPF configuration without microscales contributions $\bar{\pi}_{IJ}=0$. In this case the macroscopic contributions alone fail to account for the stress of the fluid, leading to incorrect velocities profiles. Additionally, in figure \ref{fig:newFlowCylLHMM_Ma}, we compare the velocity profiles of Newtonian fluid modelled from LHMM and full macro representations. Consistently, LHMM recovers the velocity profiles even for complex flow configurations. 

\section{Velocity profile evolution for oligomer melts}

\begin{figure}
\centering
	\includegraphics[width=1\textwidth]{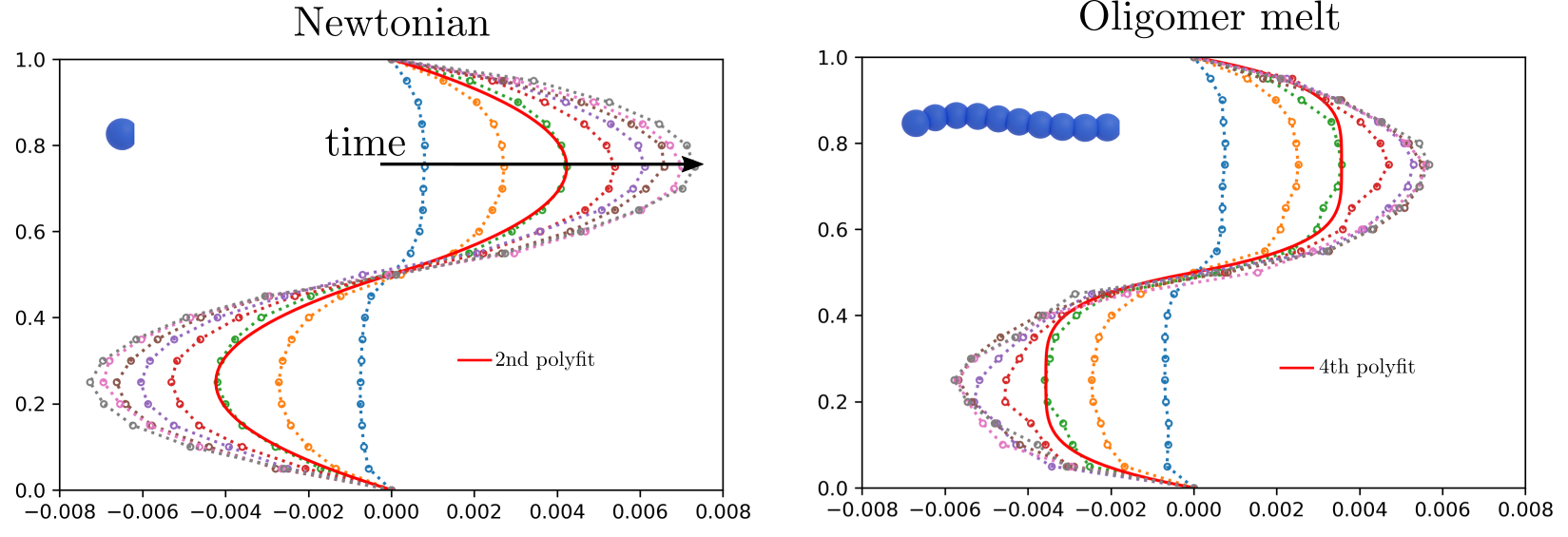}
\caption{Start-up flow in RPF configurations for  Newtonian and non-Newtonian fluids. The oligomer melt corresponds to chains with $N_s=8$. The best fitting of the velocity profile is illustrated by the continuous line at the same time step for both fluids. For Newtonian fluid is consistent with the expected quadratic profile, whereas, for oligomer melt, the microscopic effect leads to a 4th order velocity profile.}
\label{fig:startup}
\end{figure}

In addition to the steady state solution, we were able to capture the characteristic deviations in the temporal evolution for the oligomer melt. In figure \ref{fig:startup}, we compare the velocity profile stabilization for the RPF for both Newtonian and non-Newtonian fluids, under the same flow conditions. In figure \ref{fig:startup}, the solid lines correspond to the best fitting of the velocity at the same time step for both fluids. For the Newtonian fluid, the velocity profile is consistently reproduced by a quadratic fitting, whereas the microscopic effects of the oligomer chains lead to a 4th order velocity profile in the non-Newtonian fluid.

\section{Velocity and stress evolution for multiphase systems}

The evolution of the velocity and stress profiles in the square contraction  array for multiphase flows evidences that for these complex systems a fully developed steady stated has not been reached. The dynamic formation and destruction of microstrutures is responsible for the constant evolution of the stress.   
\begin{figure}
\centering
	\includegraphics[width=0.7\textwidth]{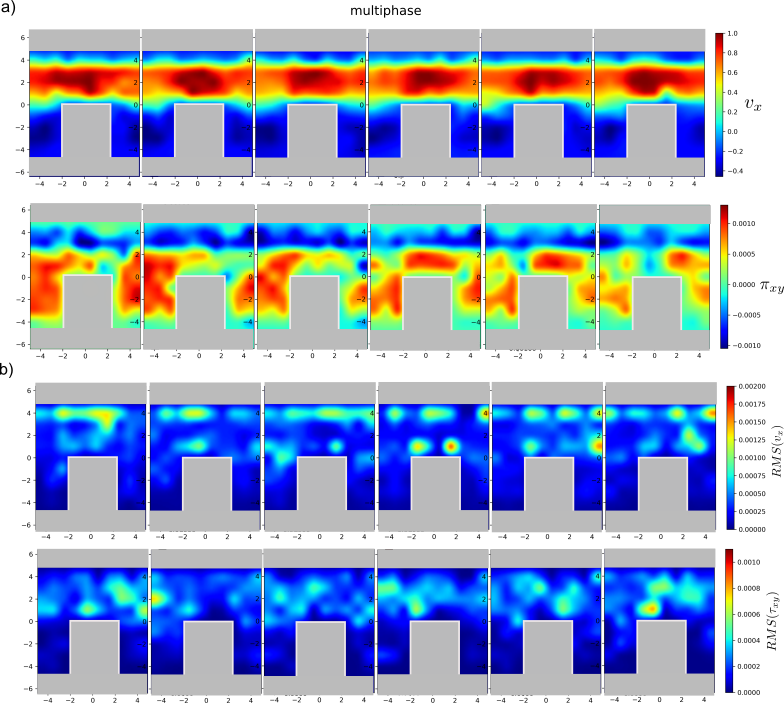}
\caption{Stabilization of velocity and stress for Newtonian and multiphase flows. a) Comparison of velocity profiles at different time steps between Newtonian and multiphase system. b) RMS of velocity and stress for multiphase systems at different time steps.}
\label{fig:mfevol}
\end{figure}

% \backsection[Author ORCID]{Authors may include the ORCID identifers as follows.  F. Smith, https://orcid.org/0000-0001-2345-6789; B. Jones, https://orcid.org/0000-0009-8765-4321}

% \backsection[Author contributions]{Authors may include details of the contributions made by each author to the manuscript, for example, ``A.G. and T.F. derived the theory and T.F. and T.D. performed the simulations.  All authors contributed equally to analysing data and reaching conclusions, and in writing the paper.''}

%\bibliographystyle{jfm}
%\bibliography{jfm}
%Use of the above commands will create a bibliography using the .bib file. Shown below is a bibliography built from individual items.

\bibliographystyle{jfm}

\bibliography{refs}

%% End of file `jfm2esam.bib'.

\end{document}